 \documentclass[final,5p,times,twocolumn,authoryear, nopreprintline]{elsarticle}

\usepackage{amssymb}
\usepackage{lipsum}
\usepackage{amsmath} 
\usepackage{booktabs} 
\usepackage[colorlinks,linkcolor=green, anchorcolor=green,citecolor=green]{hyperref}

\usepackage{pdflscape}
\usepackage{rotating}
\usepackage{makecell}
\usepackage[normalem]{ulem}

\usepackage{xcolor}

\usepackage{lineno}

\begin{document}

\begin{frontmatter}



\title{A Helical Trajectory Interpretation of a Multi-wavelength Event in a Blazar S5 0716+714}


\author[1,2]{Zhihao Ouyang\corref{cor1}}
\affiliation[1]{organization={Shanghai Key Lab for Astrophysics, Shanghai Normal University}, 
            city={Shanghai},
            postcode={200234}, 
            country={People's Republic of China}
            }
\affiliation[2]{organization={Center for Astrophysics, Guangzhou University}, 
            city={Guangzhou},
            postcode={510006}, 
            country={People's Republic of China}
            }
\cortext[cor1]{Zhihao Ouyang and Jingyu Wu contribute equally to this work.}

\author[1]{Jingyu Wu\corref{cor1}}
\author[1]{Hubing Xiao\corref{cor2}} \ead{hubing.xiao@shnu.edu.cn}
\cortext[cor2]{Corresponding author}

\author[1]{Zhijian Luo}
\author[1]{Zhu Chen}
\author[1]{Shaohua Zhang}
\author[3]{Liang Chen}
\affiliation[3]{organization={Shanghai Astronomical Observatory, Chinese Academy of Sciences}, 
            city={Shanghai},
            postcode={200030}, 
            country={People’s Republic of China}
            }

\author[1]{Jianzhen Chen}

\author[4]{Marina Manganaro}
\affiliation[4]{organization={Department of Physics, University of Rijeka}, 
            city={Rijeka},
            postcode={51000}, 
            country={Croatia}
            }

\author[5,6]{Luis C. Ho}
\affiliation[5]{organization={Kavli Institute for Astronomy and Astrophysics, Peking University}, 
            city={Beijing},
            postcode={100871}, 
            country={People's Republic of China}
            }
\affiliation[6]{organization={Department of Astronomy, School of Physics, Peking University}, 
            city={Beijing},
            postcode={100871}, 
            country={People's Republic of China}
            }
            
\author[2]{Junhui Fan}

\author[1]{Chenggang Shu}

\begin{abstract}
Relativistic jets in blazars exhibit strong variability across multiple timescales, but the connection between temporal variability and the evolution of jet structures remains poorly understood.
In this work, we investigate the $\gamma$-ray variability of the BL Lac object S5 0716+714 from July 2014 to June 2015 and explore its possible connection with the parsec-scale jet evolution.
We identify an escalating periodic oscillation (EPO) in the $\gamma$-ray light curve, characterized by a sequence of successive flux enhancements with progressively increasing temporal separations.
Combining \textit{Fermi}-LAT data with contemporaneous 43 GHz Very Long Baseline Array (VLBA) observations, we find a temporal association between the EPO and the downstream propagation of a superluminal knot.
A helical trajectory model simultaneously fitting the $\gamma$-ray variability and the two-dimensional VLBA component positions reproduces the observed flux modulation and component motion.
These results suggest that geometric Doppler modulation caused by the changing viewing angle of a moving jet component provides a possible explanation for the observed EPO.
This work provides a framework of connecting blazar variability to the knot evolution of relativistic jets.
\end{abstract}




\end{frontmatter}



\section{Introduction}\label{sec:intro1}

Blazars are one of the active galactic nuclei (AGNs) whose relativistic jets are viewed at small angles to the line of sight.
Their observed non-thermal emission is therefore strongly enhanced by Doppler beaming \citep[e.g.,][]{Urry1995, Blandford2019ARA&A57}.
They are divided into two subclasses, BL Lacertae objects (BL Lacs) and flat spectrum radio quasars (FSRQs), based on the presence of optical emission lines \citep{Stickel1991, Scarpa1997}.
Blazars are also characterized by the rapid and strong variability in multiple wavelengths, with the timescale ranging from minutes to years \citep[e.g.,][]{Ulrich1997ARA&A35, Aharonian2007ApJ, Abdo2010ApJ722, Ackermann2016ApJL, Goyal2017ApJ837, Wang2024ApJS270, Raiteri2025A&ARv33}.
Their broadband spectral energy distributions (SEDs) typically exhibit two broad humps. The low-energy component, extending from radio to optical/UV or X-ray bands, is generally attributed to synchrotron radiation from relativistic electrons in the jet.
The high-energy component is commonly interpreted as inverse Compton emission in leptonic scenarios or radiation associated with relativistic protons and secondary particles in hadronic scenarios \citep[e.g.,][]{Bottcher2013, Cerruti2015MNRAS, Tavecchio1998, Ghisellini2009MNRAS397, Gao2019NatAs, Xue2019ApJ871, Xue2025EPJC85, Ouyang2025ApJ980, Xiao2026ApJ1005}.

Very long baseline interferometry (VLBI) has been extensively used to study the parsec-scale structure and evolution of blazar jets.
Long-term VLBI monitoring has revealed that blazar jets contain both moving and quasi-stationary features and often exhibit complex kinematic evolution.
The Monitoring Of Jets in Active galactic nuclei with VLBA Experiments (MOJAVE)  program has systematically investigated the parsec-scale morphology and kinematics of large samples of radio-loud AGNs using the 15 GHz Very Long Baseline Array (VLBA) observations.
For example, \citet{Lister2009} measured the motions of hundreds of jet components and found apparent speeds extending to tens of times the speed of light, and subsequent studies showed that acceleration, non-radial motions, and structural evolution are common on parsec scales \citep[e.g.,][]{Lister2013, Lister2018, Lister2019}.
At 43 GHz, the VLBA-BU-BLAZAR monitoring program at Boston University has monitored $\gamma$-ray-bright blazars, providing detailed measurements of the kinematics and evolution of individual jet components \citep{Jorstad2005, Jorstad2017, Weaver2022}. 
Based on long-term monitoring, these studies have traced a large number of jet components and revealed a broad range of apparent speeds, as well as acceleration, non-ballistic motions, and the presence of quasi-stationary features.

Whether and how variability is connected to parsec-scale jet evolution remain open questions.
Previous studies have investigated the possible connections through high-energy activity.
Individual $\gamma$-ray flares or outbursts have been reported to be associated with either the emergence of new superluminal components or a moving component passing through a stationary component \citep[e.g.,][Ouyang et al. in prep.]{Marscher2008Natur452, Jorstad2013ApJ773, Casadio2015ApJ813, Kim2022ApJ925, MAGIC2018AA619, MAGIC2018AA617}.
However, these studies have mainly established the connections between individual flares and specific states of jet components.
It remains challenging to determine whether the long-term blazar variability can be related to the motion of an individual jet component.
Previous studies of quasi-periodic oscillations (QPOs) have attempted to investigate such a connection.
For QPO candidates with periods of tens of days, such as transient periods of 34.5 days in PKS 2247-131 and 31.3 days in S5 0716+714 \citep{Zhou2018NatCo9, Chen2022ApJ}, geometric effects associated with the motion of emitting blobs along helical trajectories in the jet have been proposed as possible explanations.
In this scenario, the variability could be related to the evolution of the jet component as it propagates downstream.
However, this interpretation is primarily constrained by the observed light curves, while the inferred geometric evolution has not been directly compared with the contemporaneous trajectory of the emitting blob.
Therefore, a close connection between variability and the contemporaneous evolution of jet components has yet to be established observationally.

In this work, we investigate whether the observed variability of the blazar S5 0716+714 and the contemporaneous motion of its parsec-scale jet component can be connected within a geometric framework.
S5~0716+714 is a highly variable BL Lacertae object with a statistically inferred redshift of $z=0.2304$ \citep{Pichel2023A&A680}.
During the period from July 2014 to June 2015, the $\gamma$-ray light curve shows five consecutive flux enhancements with both their amplitudes and successive separations increasing with time.
We refer to this behavior as an escalating periodic oscillation (EPO), distinguishing it from a typical QPO with an approximately stationary characteristic period.
Contemporaneous 43 GHz VLBA observations reveal a superluminal component propagating downstream along the parsec-scale jet during the same time interval. 
This temporal coincidence motivates us to investigate the connection between the $\gamma$-ray EPO and the motion of the jet component, and to examine whether both could be explained by a geometric origin.
To test this, we employ a simple conical helical jet model to examine whether the increasing oscillation period, the flux modulation, and the motion of the jet component can be described within a geometric framework.
This study therefore provides an exploration to connect high-energy variability with parsec-scale jet kinematics and to probe the geometric structure and dynamical evolution of the relativistic jet.
Throughout this work, we adopt the $\Lambda$CDM cosmological parameters with $H_{0} = 67.7$~km~s$^{-1}$~Mpc$^{-1}$ and $\Omega_{\rm m} = 0.310$ \citep{Planck2020A&A641}.
At $z=0.2304$, this corresponds to a luminosity distance of $d_{L}=1.186$ Gpc, with 1 mas corresponding to an angular scale of 3.799 pc.

\section{Observations and Results}\label{sec:method}

\subsection{Fermi-LAT data reduction}

\begin{figure*}
\centering
\includegraphics[width=0.85\textwidth]{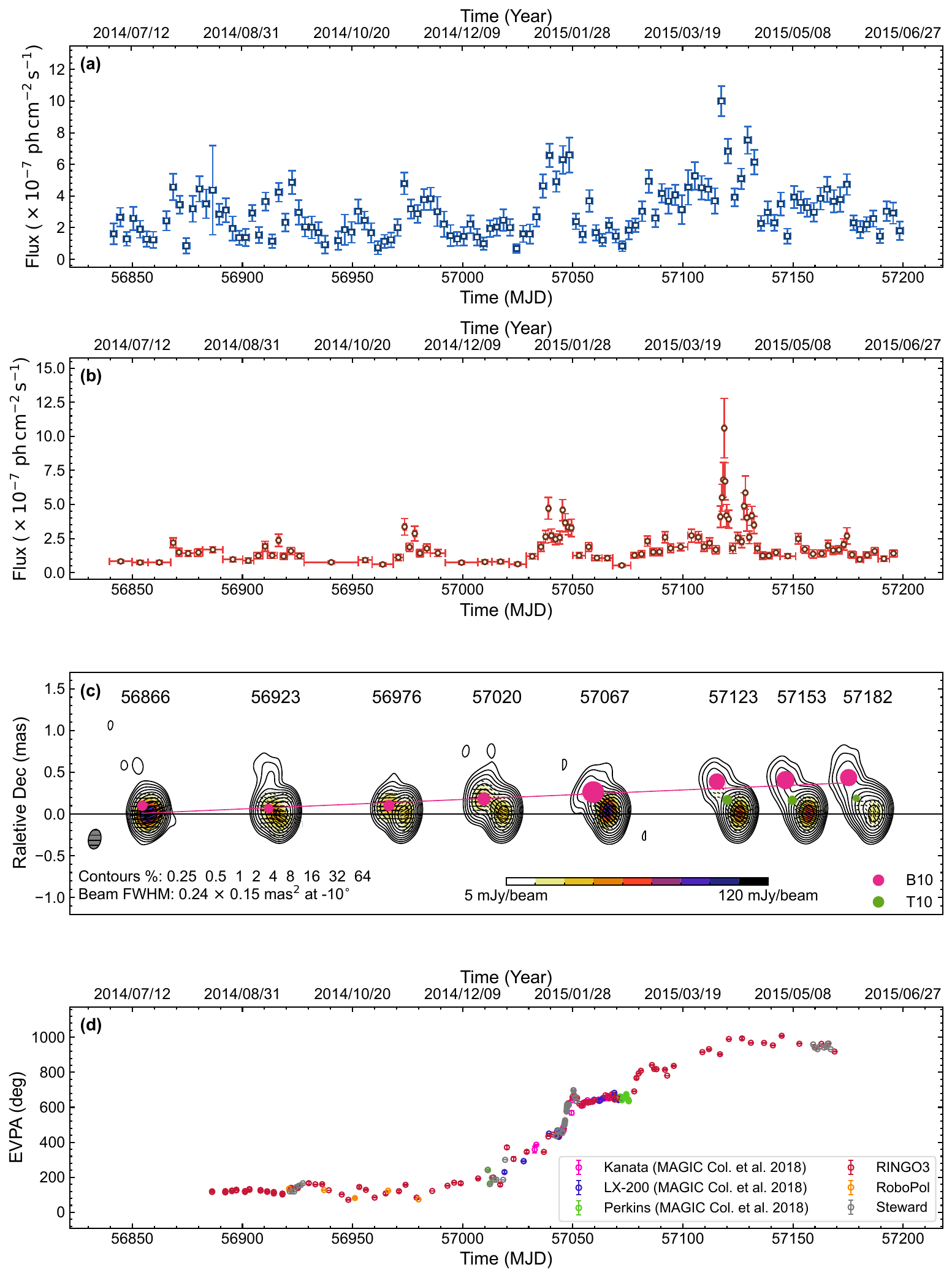}
\caption{
Panel (a): $\gamma$-ray 3-day-binned light curve of S5 0716+714 over MJD 56840--57200.
Panel (b): $\gamma$-ray adaptive-binning light curve of S5 0716+714 over MJD 56840--57200. 
Panel (c): Sequence of total-intensity (contours) and polarized-intensity (colour scale) images of S5 0716+714 at 43 GHz between MJD 56866 and 57182.
The pink and green circles mark knots B10 and T10, respectively.
The pink line shows the simple linear fit to the B10 distance relative to the core.
Black line segments in each image indicate the polarization direction.
The solid black horizontal line marks the position of the core A0.
Panel (d): Electric vector position angle (EVPA) at R band as a function of time. The error bars are relatively small compared with the marker size. 
Different colored symbols denote measurements from different instruments.
}
\label{fig:LC_VLBA_EVPA}
\end{figure*}

We collected the LAT data from the \emph{Fermi}-LAT Pass 8 database covering the entire 16-year observation in the energy range of 0.1--300 GeV.
The photon events were selected within a $15^{\circ}$ radius region of interest (ROI) centered on the location of S5 0716+714, and excluded the zenith angles larger than  $90^{\circ}$ to reduce contamination from $\gamma$-rays produced by the Earth's limb.
We performed the unbinned likelihood analysis using the latest {\ttfamily Fermitools} \citep[v2.2.0;][]{Fermitools2019ascl} and the instrument response functions (IRFs) {\ttfamily P8R3\_SOURCE\_V3} to perform light curve analysis. 
The condition ``{\ttfamily evclass=128, evtype=3}'' was used to filter events with a high probability of being photons, and ``{\ttfamily (DATA\_QUAL$\geqslant$0)\&\&(LAT\_CONFIG==1)}'' was used to select the good time intervals.
The model file, generated by package {\ttfamily LATSourceModel}\footnote{https://github.com/physicsranger/make4FGLxml}, included all the sources from the latest \emph{Fermi}-LAT Fourth Source catalog \citep[4FGL-DR4;][]{Abdollahi2022ApJS260} within 25$^{\circ}$ of the target source, as well as the Galactic ({\ttfamily gll\_iem\_v07.fits}) and extragalactic isotropic ({\ttfamily iso\_P8R3\_SOURCE\_V3\_v1.txt}) diffuse emission components.
The spectral parameters of sources with an average detection significance greater than $5\sigma$ within $5^{\circ}$ of the ROI, as well as those of the diffuse components, were set free.
Based on the best-fit source model, we generated a $\gamma$-ray light curve with 3-day time bins assuming a power-law spectral model for the source.
Only the flux points with a test statistic\footnote{The test statistic is defined as
${\rm TS} = -2 \ln \left(\frac{\mathcal{L}_{\rm max,0}}{\mathcal{L}_{\rm max,1}}\right)$, where $\mathcal{L}_{\rm max,0}$ and $\mathcal{L}_{\rm max,1}$ are the maximum likelihood values for the models without and with the source, respectively. 
A larger TS value indicates a higher significance of the source detection \citep{Mattox1996ApJ461}.} of TS$\geq$9 were included; 
otherwise, the 95\% confidence level upper limits were calculated using the {\ttfamily UpperLimits}\footnote{https://fermi.gsfc.nasa.gov/ssc/data/analysis/scitools/upper\_limits.html} tool. 
In addition, we generated an adaptive-binning $\gamma$-ray light curve following \citet{Lott2012A&A544} to track flux variations without imposing an arbitrary temporal cadence.
The adaptive time bins were optimized to achieve a constant relative flux uncertainty of 15\% above the optimal energy $E_{\rm opt}=209.62$ MeV \citep{Lott2012A&A544}. 
This adaptive-binning strategy avoids the introduction of predefined timescales and flux upper limits associated with constant-binning methods.
The resulting 3-day-binned and adaptive-binning light curves are shown in the upper two panels of Figure~\ref{fig:LC_VLBA_EVPA}.

\subsection{Identification and significance of escalating periodic oscillation}\label{sec:EPO_significance}

Figure~\ref{fig:LC_VLBA_EVPA} presents both the 3-day-binned and adaptive-binning $\gamma$-ray light curves.
These two light curves provide complementary views of the temporal variability, with adaptive binning resolving rapid flux changes during bright states and the 3-day-binned light curve independently tracing the longer-term flux modulation.
During MJD 56840--57200, the adaptive-binning light curve shows a sequence of five enhancements.
These enhancements show increasing peak fluxes and a systematic increase in the temporal separation between successive peaks, suggesting an evolution of the characteristic oscillation timescale.

\begin{figure*}
\centering
\includegraphics[width=0.8\textwidth]{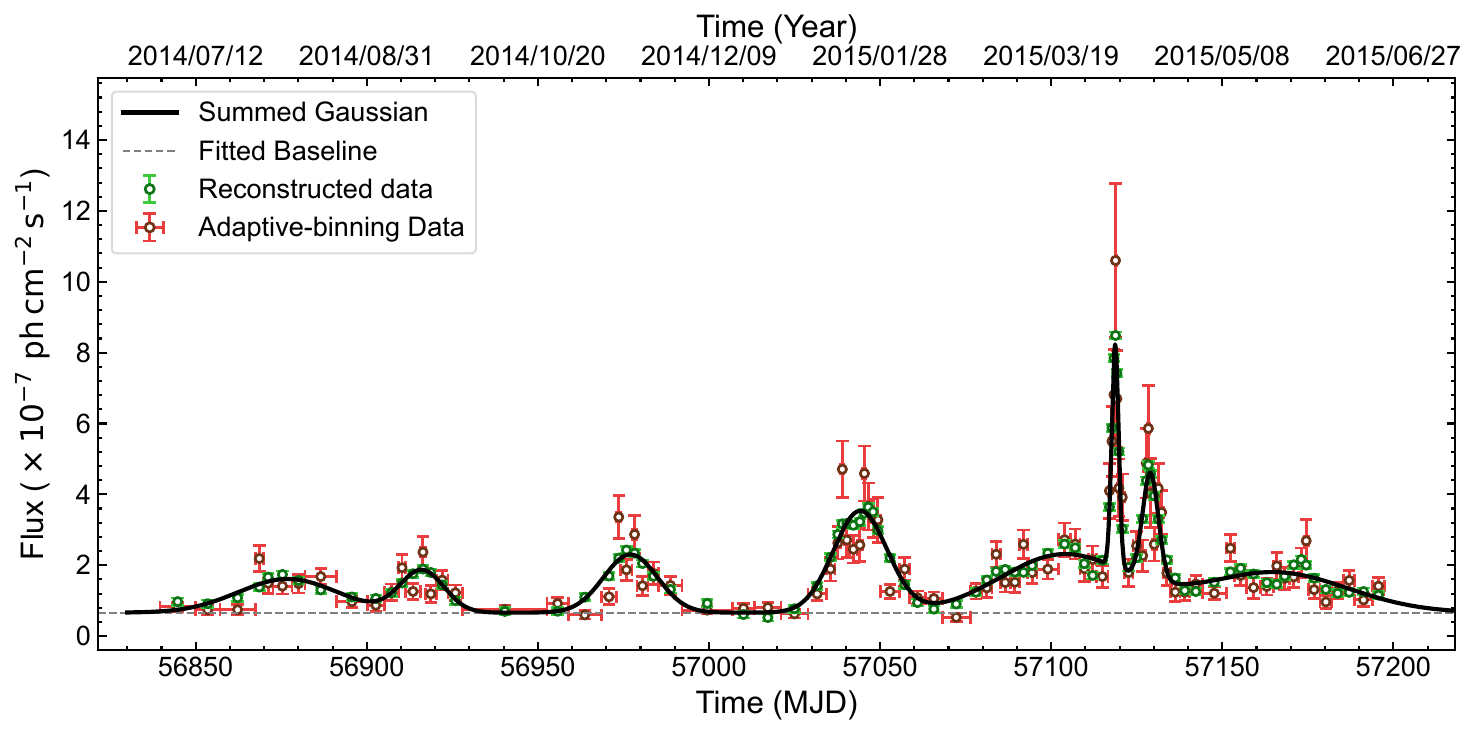}
\caption{Illustration of the escalating periodic oscillation (EPO).
The red points represent the adaptive-binning light curve, while the green ones represent the reconstructed light curve
Gaussian profiles are fitted to the reconstructed $\gamma$-ray light curve over MJD 56840--57200 to identify individual flux enhancements, while the original adaptive-binning measurements are overplotted only for visual comparison.
}
\label{fig:Gaussian_B10}
\end{figure*}

We employed a fast Fourier transform (FFT) technique to transform the adaptive-binning light curve into the frequency domain and remove high-frequency components arising from random rapid $\gamma$-ray variability \citep[e.g.,][]{Meyer2019ApJ877, Wang2024ApJS270}.
In this work, we consider only the low-frequency components with $f < 0.0456~{\rm day}^{-1}$, which is sufficient to reproduce the observed light curve.
In the end, the filtered signal was transformed back to the time domain through an inverse FFT, yielding a reconstructed $\gamma$-ray light curve with a clearly identifiable low-frequency variability pattern.
The reconstructed low-frequency light curve shows the sequence of five prominent enhancements with progressively increasing temporal separations. 
These enhancements are described by a model consisting of a constant flux component and five additional Gaussian components, and the fitting result is shown in Figure~\ref{fig:Gaussian_B10}.

\begin{figure*}
\centering
\includegraphics[width=0.8\textwidth]{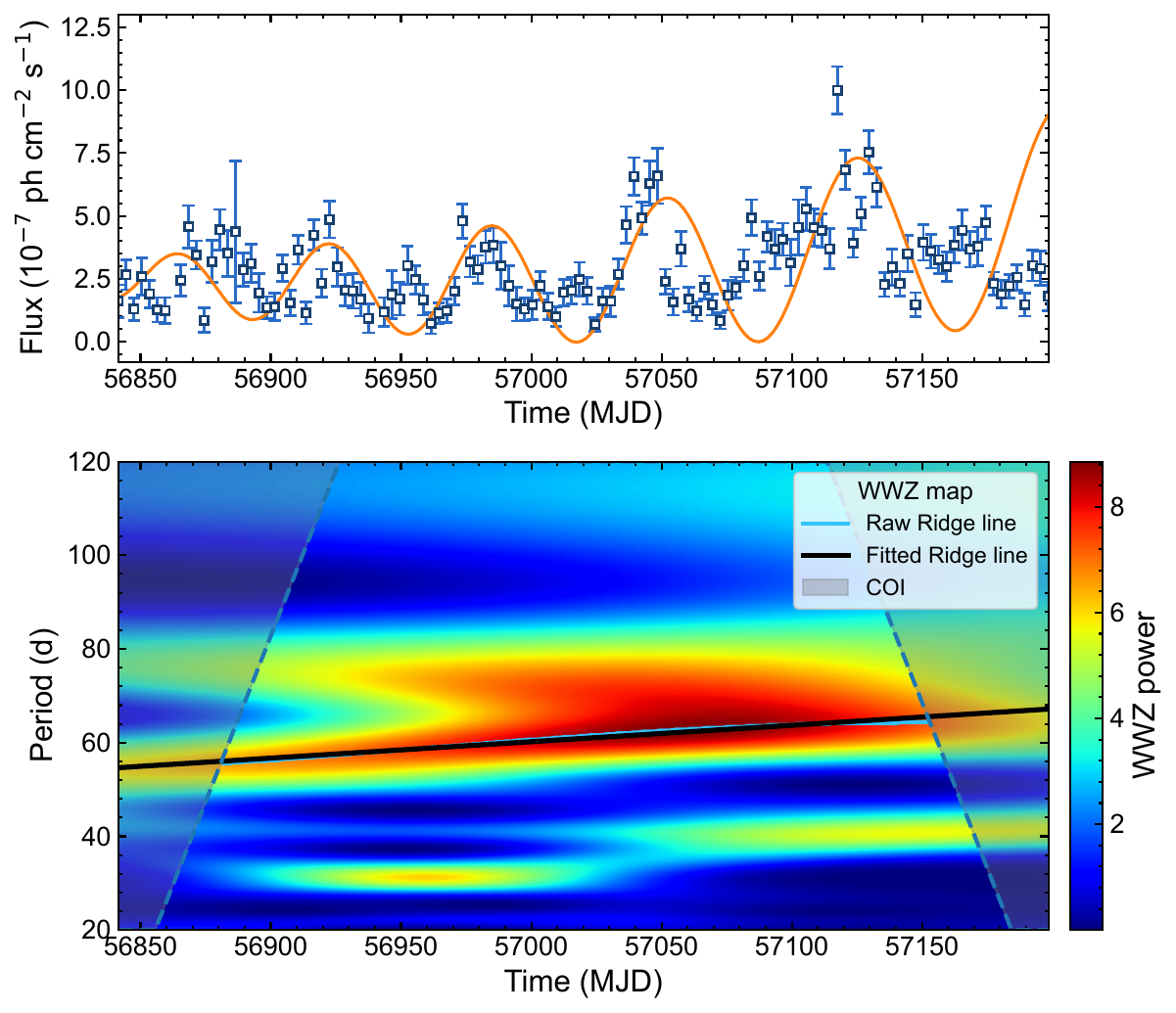}
\caption{
Illustration of the WWZ analysis. 
The upper panel shows the 3-day-binned $\gamma$-ray light curve. 
The orange line shows, for illustrative purposes only, a sinusoidal function with a period-increasing rate of $\dot{P}=0.034$ derived from the ridge linear fitting, whose amplitude also increases with period.
The lower panel displays the corresponding WWZ power map. 
The shaded region indicates the cone of influence (COI). 
The light-blue line denotes the raw ridge extracted from the periods of maximum WWZ power, and the black line shows the fitted ridge.
}
\label{fig:WWZ}
\end{figure*}

In addition, the 3-day-binned light curve was used to quantify the temporal evolution of the oscillation independently. 
We applied the weighted wavelet Z-transform \citep[WWZ;][]{Foster1996AJ112} over periods ranging from 20 to 120 days, as shown in Figure~\ref{fig:WWZ}.
At each time, the ridge period was defined as the period at which the WWZ power reaches its maximum.
Regions inside the cone of influence (COI), which account for edge effects arising from the finite length of the data, were excluded from the ridge analysis.
The resulting WWZ map reveals a continuous ridge of period evolving from approximately 54 to 67 days.
We fitted its temporal evolution using $P_{\rm ridge}(t)=P_0+\dot{P}~(t-t_0)$, and showed the result as a solid-black line in Figure~\ref{fig:WWZ}.
The period-growth rate is $\dot{P}=0.034 \pm 0.001 \, {\rm day\cdot day^{-1}}$, a positive $\dot{P}$ suggesting that the characteristic period increases with time. 
To assess whether the observed evolving ridge of period could arise from stochastic red-noise variability, we generated $N_{\rm sim}=10^{5}$ artificial light curves using the method of \citet{Emmanoulopoulos2013MNRAS433}, based on the best-fitting power spectral density (PSD) and probability density function (PDF) of the observed 3-day-binned light curve. 
The PSD and PDF fitting procedures, together with their best-fitting parameters, are presented in \ref{appendix:PSD_PDF}.
Each simulated light curve was extended to 100 times the duration of the observed interval to reduce red-noise leakage, and sampled with the same cadence as the observed light curve.
For each simulated light curve, the WWZ map was calculated using the same procedures and parameter settings as those applied to the observed light curve, followed by the same ridge extraction and linear fitting procedures.
For the local significance estimation, ridge searches were restricted to the 54--67-day period interval identified from the observed WWZ map.
The same ridge-selection criteria were applied to both the observed and simulated light curves.
A ridge was retained only if it covered at least three oscillation cycles, contained no temporal gap exceeding 0.5 local cycles, and showed no period jump larger than 5 days between adjacent ridge points.
We quantified the escalating behavior using two complementary statistics.
The primary statistic, $T_{\rm EPO}$, measures the improvement of a linearly increasing-period model ($\mathcal{M}_{1}$) relative to a constant-period model ($\mathcal{M}_{0}$).
Specifically, we compared
\begin{equation}
    \mathcal{M}_{0}:\quad P(t)=P_{0}
\end{equation}
with
\begin{equation}
    \mathcal{M}_{1}:\quad
    P(t)=P_{0}+\dot{P}(t-t_{\rm 0}),
    \qquad \dot{P}>0.
\end{equation}
Then, we defined
\begin{equation}
    T_{\rm EPO}=\max\left[0,\, {\rm RSS}(\mathcal{M}_0)-{\rm RSS}(\mathcal{M}_1)\right],
\end{equation}
where ${\rm RSS}(\mathcal{M}_{0})$ and ${\rm RSS}(\mathcal{M}_{1})$ are the residual sums of squares of the constant- and increasing-period models, respectively. 
A larger $T_{\rm EPO}$ indicates that the increasing-period model provides a better description of the data than the constant-period model.
The second statistic is the fitted period-growth rate, $\dot{P}$, which directly quantifies the strength of the period increase. 
For both statistics, a simulated candidate was considered an exceedance only if its ridge satisfied the minimum WWZ-power threshold measured from the observed data.
This ensures that the simulated ridge reached a WWZ power at least comparable to that of the observed ridge.
For each test statistic, $Q=T_{\rm EPO}$ or $\dot{P}$, the local false-alarm probability ($p$-value) was estimated as \citep{PhipsonSmyth2010}
\begin{equation}
p = \frac{k+1}{N_{\rm sim}+1},
\label{Equ:p_value_cal}
\end{equation}
with
\begin{equation}
k=\sum_{j=1}^{N_{\rm sim}}
I\left(Q_j\geq Q_{\rm obs}\right),
\end{equation}
where $Q_{\rm obs}$ is the statistic measured from the observed light curve, $Q_j$ is that obtained from the $j$-th simulation, and $I$ is the indicator function which equals 1 when the condition is satisfied and 0 otherwise.
We obtain $p_{\rm local,T_{\rm EPO}}=5.50\times10^{-4}$ and $p_{\rm local,\dot P}=5.60\times10^{-4}$, corresponding to local significances of approximately $3.26\sigma$ in both cases.

Since the 54--67-day period was selected from the observed WWZ map, we additionally evaluated the global false-alarm probability to account for the look-elsewhere effect.
We scanned the full 20--120-day period range using a sliding period window with a width of 13 days.
The window was shifted by 1 day at each step, resulting in 88 overlapping search windows.
For each simulated light curve, we retained the largest value of the test statistic among all windows satisfying the same ridge-selection and minimum WWZ-power criteria.
The global false-alarm probability was then calculated using Equation~\eqref{Equ:p_value_cal}.
For $T_{\rm EPO}$, we obtain $p_{\rm global,T_{\rm EPO}}=4.10\times10^{-3}$, corresponding to $2.64\sigma$.
Using $\dot{P}$ gives $p_{\rm global,\dot P}=4.68\times10^{-3}$, corresponding to $2.60\sigma$.
The two statistics therefore obtain consistent local and global significance estimation.
These results suggest that the observed EPO signal is unlikely to arise from the underlying red-noise process.

\subsection{Very Long Baseline Array radio data}

We analysed 43 GHz (7 mm wavelength) VLBA observations of S5 0716+714 obtained from the VLBA-BU-BLAZAR monitoring program \citep{Jorstad2016Galax4}. 
The total intensity image provided by the BU-BLAZAR team covers the period from 28 July 2014 (MJD 56866) to 9 June 2015 (MJD 57182), which had already been calibrated and self-calibrated as described in \citet{Jorstad2005, Jorstad2017}.
We subsequently modeled the total intensity image with multiple two-dimensional circular Gaussian components using the {\ttfamily modelfit} task in the {\ttfamily DIFMAP} package, following the procedures described in \citet{Jorstad2017} and \citet{Weaver2022}. 
The modeling procedure initially included a Gaussian component representing the brightest upstream feature of the parsec-scale jet. 
This feature was identified as the core component A0 and was assumed to be stationary.
Additional Gaussian components were subsequently introduced to represent other bright features in the jet.
This process was continued until the inclusion of additional components no longer resulted in a significant improvement in the $\chi^2$.
The fitted parameters, including the flux density ($S$), radial distance from the core ($r$), position angle (\textit{PA}) with respect to the core, angular size ($a$) corresponding to the full width at half maximum (FWHM), and two-dimensional position with respect to the core of each component, are listed in Table~\ref{Tab:VLBA_knot_params} and the the corresponding VLBA maps are shown in panel (c) of Figure~\ref{fig:LC_VLBA_EVPA}. 
The observed brightness temperature was then calculated as $T_{\rm b,obs} = 7.5 \times 10^{8} S / a^{2}$~K.
The uncertainties in the fitted parameters were estimated following the method described by \citet{Weaver2022}.
Except for the brightest core A0, all other Gaussian components were referred to as ``knots''.
The knots were labeled according to their order of appearance after the beginning of the VLBA monitoring program. 
The knots were identified across epochs by requiring that the fitted flux density, radial distance, position angle, and angular size of the components should not change abruptly with time given the regularity of our observations.
Our fitted component parameters are consistent with those reported in previous VLBA monitoring; therefore, we adopted the knot nomenclature used in their work. 
Accordingly, we identified two stationary knots, A1 and A2, and two moving knots, B10 and T10.
Among the moving knots, B10 exhibits a notable kinematic evolution during the period of interest. 
This evolution occurs contemporaneously with the $\gamma$-ray EPO signal, suggesting a possible connection between the propagation of the radio knot and the observed $\gamma$-ray variability.

We further analyzed the kinematics of knot B10 using the fitted parameters to derive its kinematic properties under a redshift of $z=0.2304$.
Specifically, we followed the procedure reported in Section 2.2 of \citet{Weaver2022} and calculated the apparent velocity and the ejection time (i.e., the time at which the ejection passed through the core, derived from the extrapolation of motion) from the piecewise fits.
The resulting kinematic parameters are as follows.
The derived proper motion corresponds to an apparent velocity of $\beta_{\rm app, VLBA}=7.239\pm1.186$. 
The ejection time of B10 was estimated by extrapolating the fitted motion back to the core position and was found to be MJD $56761.96\pm42.04$.

\subsection{Optical polarization data}\label{sec:method_polarization}
We collected optical $R$ band polarization measurements from the Steward Observatory spectropolarimetric monitoring program \citep{Smith2009arXiv0912}, which provides long-term optical spectropolarimetry of blazars, and from the RoboPol project \citep{Blinov2021MNRAS501}, an optical polarization monitoring program for active galactic nuclei. 
Additional polarization data were adopted from \cite{MAGIC2018AA619}, including observations obtained with the 1.5-m Kanata, 40-cm LX-200, and 1.83-m Perkins telescopes.

In addition, we analysed optical polarimetric observations acquired with the RINGO3 instrument on the Liverpool Telescope\footnote{https://telescope.ljmu.ac.uk/TelInst/Inst/RINGO3/}.
RINGO3 is equipped with three cameras, namely `Red', `Green', and `Blue'.
We only used the data from the `Green' camera, as its effective wavelength is closest to that of the $R$ band.
The data reduction and polarization analysis followed the procedures described in \cite{Maund2021MNRAS503}.

The electric vector position angle (EVPA) of polarization is measured modulo 180$^{\circ}$ and is conventionally reported within the interval $[0^{\circ},~180^{\circ})$, resulting in an intrinsic $n\pi$ ambiguity.
This means that EVPA values differing by an integer multiple of 180$^{\circ}$ are mapped to the same interval $[0^{\circ},~180^{\circ})$ and are therefore observationally indistinguishable.
To remove this ambiguity and visualize a continuous EVPA time series, we applied an unwrapping procedure following \citet{Kiehlmann2016A&A590}. 
Specifically, for two adjacent data points, the offset between them was calculated through 
\begin{equation}
    \Delta \chi_{\rm red}=|\chi_{i}-\chi_{i-1}| - \sqrt{e^{2}_{\chi_{i}} + e^{2}_{\chi_{i-1}}},
\end{equation}
where $\chi_{i}$, $e_{\chi_{i}}$, $\chi_{i-1}$, and $e_{\chi_{i-1}}$ are the two adjacent EVPA measurements and the corresponding uncertainties.
If the offset is greater than 90$^{\circ}$, the $n\pi$ were added to or subtracted from $\chi_{i}$ until the smallest possible value of $\Delta \chi_{\rm red}$ was obtained.
The resulting EVPA time series shows a prolonged rotation between July 2014 and June 2015, with a cumulative rotation of approximately $900^{\circ}$, as shown in the lower panel of Figure~\ref{fig:LC_VLBA_EVPA}.
This EVPA evolution overlaps with a substantial portion of the $\gamma$-ray EPO episode and the contemporaneous VLBA monitoring of knot B10, suggesting that the polarization evolution may be related to changes in the jet geometry during this period.

\section{The helical jet model}\label{sec:method_model}

Contemporaneous 43-GHz VLBA observations resolve the downstream motion of knot B10 toward the northeast during the interval in which the EPO is observed.
The kinematic evolution of B10 is resolved in eight epochs during this period (panel (c) of Figure~\ref{fig:LC_VLBA_EVPA}).
The temporal coincidence between the EPO and the downstream propagation of B10 suggests that the observed $\gamma$-ray variability may associate with the motion of this component along the jet.
The ejection epoch of B10 is estimated to be MJD $56761.96\pm42.04$, indicating that the knot had already been propagating downstream before the $\gamma$-ray EPO observed during MJD 56840--57200.
At its first VLBA detection, B10 is located $0.163\pm0.027$ mas from the core, approximately beyond the innermost $\sim$0.1~mas region where significant jet bending has previously been reported \citep{Kravchenko2020ApJ893}. 
The jet axis therefore appears approximately straight on the spatial scales traced by B10.
Considering the spatial and temporal association between the downstream motion of B10 and the observed EPO signal, we therefore test whether a trajectory can simultaneously reproduce the two-dimensional VLBA positions of B10 and the $\gamma$-ray variability.

\begin{figure*}
\centering
\includegraphics[width=0.65\textwidth]{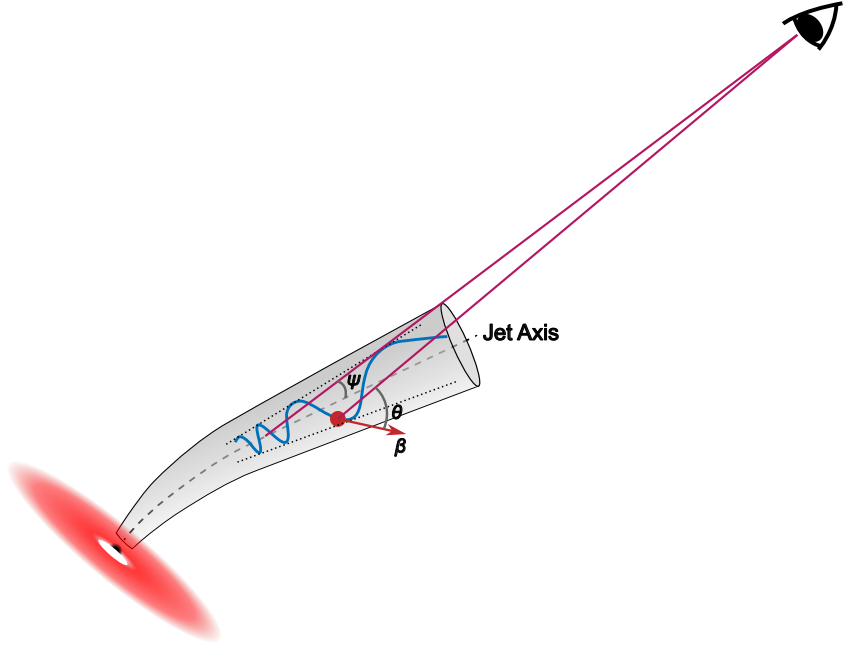}
\caption{
Illustration of the helical trajectory model of a moving blob.
Note that the dimensions in this schematic are not drawn to scale. 
A bent structure is observed in the inner part of the S5 0716+714 jet, whereas the outer part appears as a straight extension, as reported by \cite{Kravchenko2020ApJ893}. 
In this diagram, the black solid circle represents the central black hole, and the surrounding red region indicates the accretion disk. 
The gray region corresponds to the relativistic jet, while the dashed black line denotes the jet axis. 
The blob, depicted as a red solid circle, is assumed to move along a blue helical trajectory within the straight section of the jet. 
The red arrow indicates the blob velocity ($\beta$).  
The angle $\psi$ denotes the inclination angle of the jet axis to the line of sight, while $\theta$ represents the viewing angle between the line of sight and the blob’s direction of motion.
}
\label{fig:model}
\end{figure*}

We model this behavior by treating the knot B10 as an idealized emitting blob.
We therefore adopt the helical kinematic model described as Case 3 in \citet{Steffen1995A&A302}, in which the blob travels with relativistic velocity $\boldsymbol{\beta}=\boldsymbol{v}/c$ along a helical trajectory on the surface of a cone with half-opening angle $\Omega$.
In the jet's rest frame, the jet axis is aligned with the $z$-axis of a Cartesian coordinate system.
The observer's line of sight is characterized by an inclination angle $\psi$ to the jet axis and an azimuth angle $\varepsilon$, as shown in Figure~\ref{fig:model}. 
The model assumes conservation of kinetic energy, angular momentum (along the jet axis), and half-opening angle:
\begin{itemize}
  \item Conservation of kinetic energy :
    $E_{\rm kin} = \Gamma m_{0} c^{2} = {\rm constant}$
  \item Conservation of angular momentum:
    $L = \omega(t) r(t)^{2} = \omega_{0} r_{0}^{2}={\rm constant}$
  \item Conservation of opening angle:
    $\tan \Omega = {\rm constant}$
\end{itemize}
where $\Gamma$ is the bulk Lorentz factor of the blob, $m_0$ is the rest mass of the blob, and $\omega_0$ and $r_0$ denote the initial angular velocity and radius, respectively.
We solve the equations of motion analytically in Cylindrical coordinates ($r$, $\phi$, $z$), assuming the VLBI core of S5 0716+714 as the coordinate origin \citep{Steffen1995A&A302}:
\begin{equation}
    r(t) = \sqrt{(v t \sin \Omega )^{2} + 2 v r_{0}  \sqrt{1-\frac{(\omega_0 r_0)^2}{v^2} }\sin \Omega +r_0^2} ,
\end{equation}
\begin{equation}
    \phi(t) = \phi_{0} + \frac{1}{\sin \Omega } \left[ 
    \arctan \frac{v^{2} t  \sin \Omega  + \sqrt{v^{2} r_{0}^{2} - {L^{2}}}}{L} - 
    \arctan \frac{\sqrt{v^{2} r_{0}^{2} - {L^{2}}}}{L}
    \right],
\end{equation}
\begin{equation}
    z(t) = \frac{r(t) - r_{0}}{\tan \Omega } ,
\end{equation}
where $\phi_{0}$ is the initial phase and $t$ refers to the time interval in the jet comoving frame.
The blob velocity components in Cartesian coordinates are expressed by:
\begin{align}
    v_x &= \sqrt{v^{2} - v_{\phi}^{2}}\sin \Omega  \cos \phi(t) - \frac{L}{r(t)} \sin \phi(t), \\
    v_y &= \sqrt{v^{2} - v_{\phi}^{2}}\sin \Omega  \sin \phi(t) + \frac{L}{r(t)} \cos \phi(t), \\
    v_z &= \frac{\sqrt{v^{2} - v_{\phi}^{2}}\sin \Omega }{\tan \Omega }.
\end{align}
The time-dependent viewing angle $\theta(t)$ between the blob velocity and the line of sight is calculated via:
\begin{equation}
    \cos \theta(t) = \frac{ \boldsymbol {\hat n} \cdot \boldsymbol {v}}{ |\boldsymbol {\hat n}| |\boldsymbol {v}| } = \frac{v_{x} \sin \psi \cos \varepsilon + v_{y} \sin \psi \sin \varepsilon + v_{z} \cos \psi}{v} {\rm ,}
    \label{Eq:viewing}
\end{equation}
where $\boldsymbol {\hat n} = \left( \sin \psi \cos \varepsilon, \, \sin \psi \sin \varepsilon, \, \cos \psi \right)$ denotes the direction of the line of sight.
This variation of the instantaneous viewing angle results in a time-dependent Doppler factor $\delta(t)$, given by
\begin{equation}
    \delta(t)=\left\{\Gamma\left[1-\beta\cos\theta(t)\right]\right\}^{-1}.
\end{equation}
Within this model, the observed $\gamma$-ray EPO signal is assumed to arise from the time-dependent Doppler factor, such that the observed flux increases as the viewing angle decreases and the Doppler factor increases.
This relation could be characterized as
\begin{equation}
    F(t) = F_{\rm int} \times \delta(t)^{p(t)} + B(t),
    \label{eq:flux_modulation}
\end{equation}
where $F(t)$ is the observed flux, $F_{\rm int}$ is the intrinsic flux in the jet comoving frame, and $B(t)$ is the baseline component accounting for the variability not captured by the helical model.
The exponent $p(t)$ depends on the emission mechanism. 
The source is classified as a BL Lac object \citep{Biermann1981ApJ247L}, for which the high-energy $\gamma$-ray emission is commonly attributed to the synchrotron self-Compton (SSC) process \citep[e.g.,][]{Bloom1996ApJ461, Mastichiadis1997A&A320}. 
We therefore adopt the SSC scenario, with $p(t)=3+\alpha$, where $\alpha$ is the spectral index defined by $F_{\nu}\propto\nu^{-\alpha}$ \citep[e.g.,][]{Dermer1995ApJ446L, Marscher2009arXiv09092576}.
Using the relation $\alpha=\Gamma_{\rm ph,\gamma}-1$, we obtain
\begin{equation}
p(t)=2+\Gamma_{\rm ph,\gamma}(t),
\end{equation}
where $\Gamma_{\rm ph,\gamma}(t)$ denotes the smooth photon index of the $\gamma$-ray band, derived from a spline interpolation of the measured photon indices.
Given the relativistic motion of the emitting blob, the observed time interval, $\mathrm{d}t_{\rm obs}$, would be much shorter than the source time, $\mathrm{d}t_{\rm src}$, in the jet comoving frame due to the relativistic beaming effect.
Therefore, during the fitting, the transformation relation between these two time intervals is characterized by \citep{Rieger2004ApJ615L}
\begin{equation}
    \mathrm{d} t_{\rm obs} = \left( 1+z \right) \left[ 1 - \beta \cos \theta(t) \right]  \mathrm{d}t_{\rm src},
\end{equation}
where $\theta(t)$ is the viewing angle derived from Equation~\eqref{Eq:viewing}. 
This relation is used to transform the model evolution in the jet comoving frame to the observer's time before comparison with the $\gamma$-ray and VLBA measurements.

The model trajectory is transformed from three-dimensional space onto the sky plane (i.e., the observational coordinate plane defined by right ascension, RA, and declination, Dec) before being compared with the VLBA measurements.
An arbitrary spatial offset $(x_0,y_0,z_0)$ is introduced to shift the conical helical trajectory, while the choice of the additive constant in $z(t)$ does not affect its projected trajectory.
The RA coordinate is sign-reversed to follow the astronomical convention that RA increases from right to left on the sky plane.
The model RA and Dec positions are fitted simultaneously to the measured relative positions of B10, together with the model fit to the observed $\gamma$-ray flux variability.
To quantify the agreement between the model predictions and the observations, we adopt the joint log-likelihood
\begin{equation}
    \ln\mathcal{L}(\Theta) = \ln\mathcal{L}_{\gamma} + \ln\mathcal{L}_{\rm RA} + \ln\mathcal{L}_{\rm Dec},
    \label{eq:joint_likelihood}
\end{equation}
where $\Theta$ denotes the set of model parameters, and each contribution is given by
\begin{equation}
    \ln\mathcal{L}_{q} = -\frac{1}{2}\sum_i \left[ \frac{\left(y_{q,i}-m_{q,i}\right)^2}{\sigma_{q,i}^2} + \ln\left(2\pi \sigma_{q,i}^2\right) \right],
    \label{eq:likelihood_individual}
\end{equation}
with $q=\{\gamma,{\rm RA},{\rm Dec}\}$ denoting the observed $\gamma$-ray flux and the relative RA and Dec positions of B10, respectively.
Here, $m_{q, i}$, $y_{q, i}$, and $\sigma_{q, i}$ are the model prediction, observed value, and associated measurement uncertainty, respectively.

We first determine the best-fitting model parameters by minimizing the negative joint log-likelihood using the {\ttfamily iminuit} package \citep{Dembinski2022zndo_iminuit}.
The fitting bounds adopted during the minimization are listed in Table~\ref{tab:model_priors_result}.
To determine the appropriate functional form of the baseline component $B(t)$ in Equation~\eqref{eq:flux_modulation}, we consider four forms: a constant ($b_{0}$), a linear function ($b_{0}+b_{1}t$), a quadratic polynomial ($b_{0}+b_{1}t+b_{2}t^{2}$), and a third-order polynomial ($b_{0}+b_{1}t+b_{2}t^{2}+b_{3}t^{3}$).
For each form, the model is fitted jointly using the likelihood defined above.
We then compare the four baseline models using the Akaike information criterion (AIC),
\begin{equation}
    {\rm AIC}=2N_{\rm free}-2\ln\mathcal{L}_{\rm max},
    \label{eq:aic}
\end{equation}
where $N_{\rm free}$ is the number of free parameters and $\mathcal{L}_{\rm max}$ is the maximum likelihood value for each baseline model.
Among these four forms, the quadratic polynomial yields the lowest AIC value.
We therefore adopt the quadratic polynomial form for $B(t)$ in the subsequent analysis.
For this adopted model, the set of free model parameters is
\begin{equation}
    \Theta = \{\Gamma,\,\psi,\,\Omega,\,r_0,\,\omega_0,\,\phi_0,\,F_{\rm int}, \,x_0,\,y_0,\,z_0,\,\varepsilon,\,b_0,\,b_1,\,b_2 \}.
\end{equation}
The resulting best-fitting parameters are subsequently used as the initial values for the Bayesian Markov Chain Monte Carlo (MCMC) analysis.

\begin{figure*}
\centering
\includegraphics[width=0.82\textwidth]{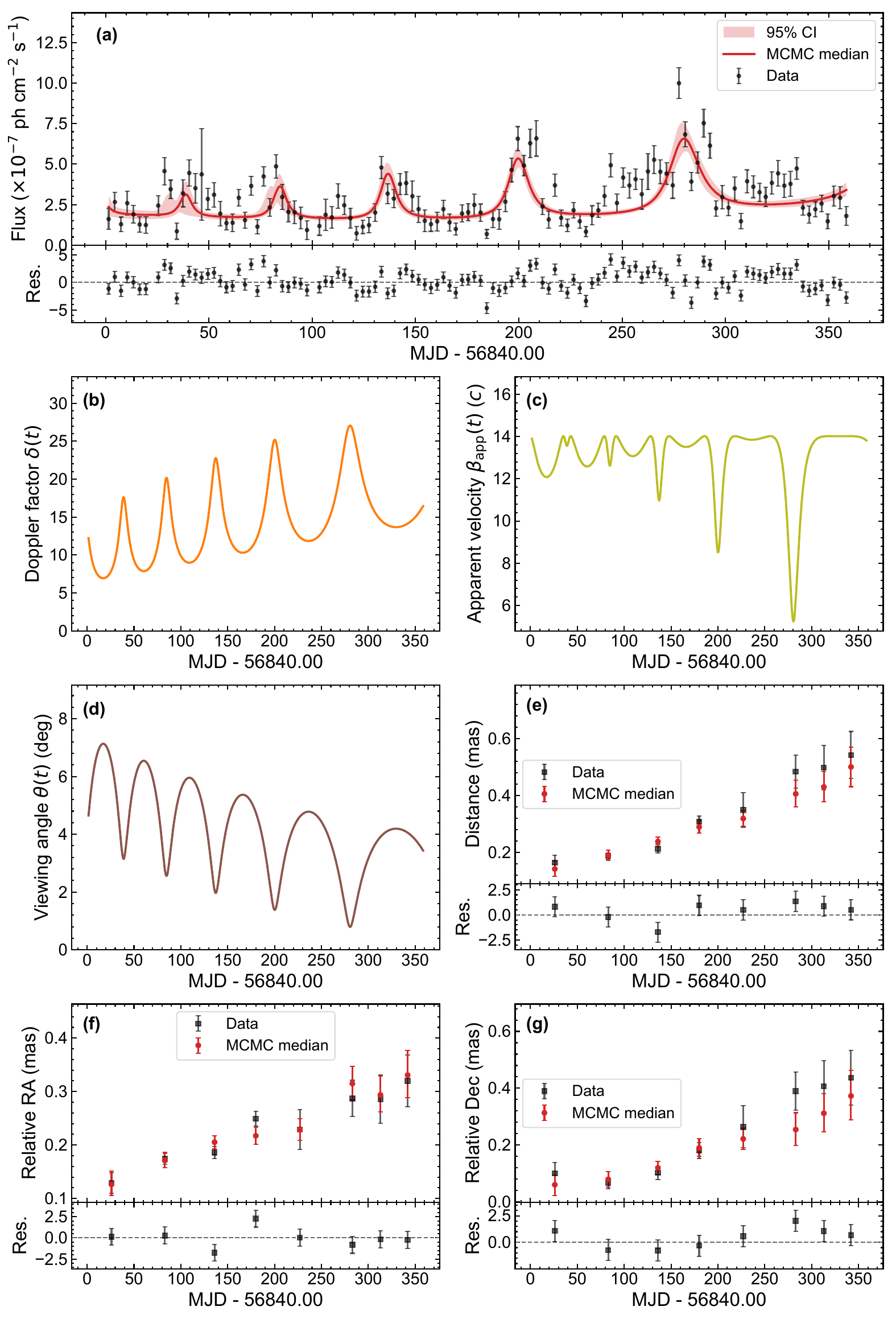}
\caption{
Simultaneous fits to the $\gamma$-ray EPO signal and VLBA position measurements of knot B10.
Panel (a) shows the $\gamma$-ray light curve with the MCMC median model and 95\% confidence interval.
The lower subpanel shows the residuals between the observations and the model.
Panels (b)--(d) show the model predicted evolution of the Doppler factor $\delta(t)$, apparent velocity $\beta_{\rm app}(t)$, and viewing angle $\theta(t)$, respectively.
Panels (e)--(g) compare the VLBA measurements with the MCMC median predictions for the radial distance, relative RA, and relative Dec, respectively.
Residuals are shown in the lower subpanels.
}
\label{fig:fitting_result_1}
\end{figure*}

We employ MCMC sampling using the {\ttfamily emcee} package \citep{Foreman-Mackey2013PASP125} to explore the posterior probability
distributions of the model parameters and estimate their uncertainties.
According to Bayes' theorem, the posterior probability, $p(\Theta|\mathrm{data})$, is proportional to the product of the likelihood and the prior,
\begin{equation}
    p(\Theta|\mathrm{data}) \propto \mathcal{L}(\Theta)\,p(\Theta).
\end{equation}
For all parameters, we adopt uniform priors within the ranges listed in Table~\ref{tab:model_priors_result}, defined as
\begin{equation}
    \ln p(\Theta) = \begin{cases}
    0, & \Theta_j^{\rm min} \le \Theta_j \le \Theta_j^{\rm max} \ \mathrm{for\ all}\ j,\\
    -\infty, & \mathrm{otherwise}.
    \end{cases}
\end{equation}
Here, $\Theta_j^{\rm min}$ and $\Theta_j^{\rm max}$ denote the lower and upper bounds of the corresponding parameter.
Posterior sampling was performed with 112 walkers for $3 \times 10^5$ iterations, with the first $3 \times 10^4$ iterations discarded as burn-in.
The convergence of the MCMC sampling was examined using the trace plots and the corresponding posterior distributions, as shown in Figure~\ref{fig:mcmc_convergence} in Appendix~\ref{appendix:mcmc_convergence}.
After the burn-in phase, the parameter chains and median traces fluctuate around stable values without systematic trends, indicating that the walkers have reached a stationary sampling region and that the MCMC sampling has adequately converged.
The medians of the marginalized posterior distributions were adopted as the parameter estimates, with uncertainties determined from the 16th and 84th percentiles.
The resulting model reproduces both the observed $\gamma$-ray flux variability and the VLBA positions of knot B10, as shown in Figure~\ref{fig:fitting_result_1}.
The corresponding normalized residuals, together with the model-derived instantaneous Doppler factor, viewing angle, and apparent velocity, are also presented in Figure~\ref{fig:fitting_result_1}.
The inferred model parameters are summarized in Table~\ref{tab:model_priors_result}, while their posterior distributions are shown in Figure~\ref{fig:mcmc} in Appendix~\ref{appendix:mcmc_convergence}.

\begin{table*}
\footnotesize
\centering
\caption{Best-fit model parameters from the MCMC analysis and parameter bounds used in the {\ttfamily iminuit} optimization and as uniform priors in the MCMC analysis. 
The best-fit values are given by the 50th percentiles of the marginalized posterior distributions, with uncertainties corresponding to the 16th and 84th percentiles, where the marginalized posterior distributions are shown in Figure~\ref{fig:mcmc}. 
Parameter $b_3$ is excluded from the MCMC analysis as mentioned in Section~\ref{sec:method_model}.}
\label{tab:model_priors_result}
\renewcommand{\arraystretch}{1.4}
\setlength{\tabcolsep}{6.7pt}
\begin{tabular}{r p{6.3cm} l l l}
\toprule
Parameters & Descriptions & Bounds & Units & MCMC Best-fit values\\
\midrule
$\Gamma$ & Bulk Lorentz factor & $[1,30]$ & -- & $14.05_{-2.03}^{+2.13}$ \\
$\psi$ & Inclination angle of the jet axis to the line of sight & $[0,10]$ & deg & $1.85_{-0.32}^{+0.41}$ \\
$\Omega$ & Half-opening angle of the conical helical trajectory & $[0,10]$ & deg & $0.09_{-0.01}^{+0.02}$ \\
$\log r_0$ & Logarithms of the initial radius & $[-2,0]$ & parsec & $-1.27_{-0.07}^{+0.07}$ \\
$\log \omega_0$ & Logarithms of the initial angular velocity & $[-6,-1]$ & rad/day & $-2.83_{-0.12}^{+0.13}$ \\
$\phi_0$ & Initial phase & $[-360, 360]$ & deg & $252.54_{-21.12}^{+26.07}$ \\
$\log F_{\rm int}$ & Logarithms of the intrinsic flux normalization & $[-15,-6]$ & ph cm$^{-2}$ s$^{-1}$ & $-11.94_{-0.23}^{+0.25}$  \\
$x_0$, $y_0$, $z_0$ & Arbitrary spatial offsets of conical helical trajectory & $[-10,10]$ & parsec & $0.19_{-0.18}^{+0.18}$, $-0.73_{-0.28}^{+0.28}$, $-0.05_{-6.80}^{+6.82}$ \\
$\varepsilon$ & Azimuth angle of the line of sight & $[-360,360]$ & deg & $-53.39_{-5.43}^{+6.13}$ \\
$\log b_0$ & Logarithms of the constant baseline coefficient & $[-10,-6]$ & -- & $-6.80_{-0.03}^{+0.03}$ \\
\parbox[t]{2.0cm}{\makebox[0.9cm][l]{$\log b_1$,} $\log b_2$,\\ \makebox[0.9cm][l]{} $\log b_3$} & Logarithms of the linear, quadratic, and third-order polynomial baseline coefficients & $[-8,-6]$ & -- & \parbox[t]{4.0cm}{$\log b_1=-7.65_{-0.21}^{+0.19}$,\\[1.5pt] $\log b_2=-7.26_{-0.18}^{+0.13}$} \\
\bottomrule
\end{tabular}
\end{table*}

\section{Discussion and Conclusion}\label{sec:discussion}

\subsection{A helical trajectory interpretation of the EPO}

During the {\emph Fermi}-LAT light-curve analysis, we found that S5~0716+714 shows five consecutive $\gamma$-ray flux enhancements with progressively increasing amplitudes and temporal separation between July 2014 and June 2015. 
Consistent with this behavior, the WWZ analysis reveals a ridge whose characteristic period increases from approximately 54 to 67 days.
The statistical local significance of this ridge reaches $3.26\sigma$ and the global significance of $2.60-2.64\sigma$, providing further evidence for the presence of the EPO signal.
To explore the possible physical origin of the EPO, we further examined the contemporaneous VLBA observations and the kinematic evolution of the parsec-scale jet.
The observed $\gamma$-ray EPO may be related to the downstream propagation of the superluminal knot B10.
The estimated ejection epoch of B10 is MJD $56761.96\pm42.04$, indicating that B10 had already been propagating downstream by the time the $\gamma$-ray EPO began.
Therefore, the EPO is unlikely to be associated with the ejection of B10 through the core, but may instead be related to its subsequent downstream propagation.
As B10 follows a helical trajectory along the conical jet, the angle between its velocity and the line of sight changes during its propagation.
These changes in the viewing angle modulate the Doppler factor and consequently produce the observed $\gamma$-ray enhancements.
If the minimum viewing angle reached during successive turns decreases, the corresponding Doppler boosting becomes progressively stronger and could account for the increasing amplitudes of the successive $\gamma$-ray peaks.
A progressive decrease in the viewing angle would enhance the Doppler boosting and could therefore account for the increasing amplitudes of the successive $\gamma$-ray peaks.
The final $\gamma$-ray peak appears to have a somewhat different profile from the preceding ones, as shown in the upper two panels of Figure~\ref{fig:LC_VLBA_EVPA}.
Around the time of this peak, B10 was located at $0.484\pm0.058$~mas from the core at MJD~57123, close to the stationary feature A2 at $\sim0.53$~mas.
At approximately the same epoch, knot T10 was first detected at a core separation of $0.235\pm0.041$~mas.
The final $\gamma$-ray enhancement may therefore include additional emission associated with T10 and/or with an interaction between B10 and the stationary feature.
Nevertheless, both scenarios retain a contribution from the downstream propagation of B10. 
We therefore consider the final enhancement to be primarily associated with B10 and include it in the model fitting used to interpret the observed EPO.
We emphasize that EPO differs from a standard QPO since its characteristic timescale of interval changes with time.
Such behavior can arise naturally if an emitting blob propagates along a helical trajectory within an expanding jet.
As the effective radius of the helical trajectory increases during downstream propagation, the viewing angle and Doppler factor change, resulting in the modulation timescale increasing systematically.
Motivated by this physical interpretation, we adopt the helical trajectory framework of \citet{Steffen1995A&A302}, as described in Section~\ref{sec:method_model}.
The model-predicted $\gamma$-ray light curve is consistent with the observed EPO (Figure~\ref{fig:fitting_result_1}, upper panel), while the predicted two-dimensional positions of B10 are consistent with the VLBA measurements (panels (e), (f), and (g) of Figure~\ref{fig:fitting_result_1}).
The model also predicts the corresponding evolution of the viewing angle, Doppler factor, and apparent velocity.
The simultaneous agreement with the $\gamma$-ray variability and the VLBA kinematics suggests that the helical trajectory of B10 provides a possible geometric interpretation of the observed EPO.
In addition, optical polarization observations during this interval reveal a continuous EVPA rotation, with a cumulative change of $\sim900^{\circ}$.
This rotation covers only part of the analyzed interval, as shown in the lower panel of Figure~\ref{fig:LC_VLBA_EVPA}, and its reconstructed evolution is sensitive to the observational cadence.
In particular, sparsely sampled intervals may miss intermediate EVPA changes \citep{Kiehlmann2016A&A590}, implying that the measured $\sim900^{\circ}$ rotation could represent a lower limit to the total rotation during the event.
The available EVPA sequence therefore does not trace the full modeled trajectory or demonstrate that the polarized optical emission and the $\gamma$-ray modulation originate from the same emission region.
Nevertheless, its temporal overlap with part of the EPO and the VLBA evolution makes it auxiliary evidence for a geometric interpretation.
In this context, the observed EVPA rotation is consistent with the propagation of a moving jet feature through a region containing a helical magnetic field, similar to the scenario proposed for BL Lacertae \citep{Marscher2008Natur452}.

Other mechanisms may nevertheless contribute to the observed variability. 
However, the systematic increase in the characteristic timescale of the EPO is not naturally explained by models that assume an approximately stationary underlying periodicity.
Binary supermassive black hole systems \citep[e.g.,][]{Villata1998MNRAS, Qian2007ChJAA} and jet precession or Lense-Thirring precession \citep{Lense1918PhyZ, Wu2025MN} can produce periodic or quasi-periodic modulation, but do not naturally explain the systematic increase of the EPO timescale.
Moreover, since the accretion disk itself cannot produce the observed high-energy $\gamma$-ray emission, which instead originates from dissipation in the relativistic jet, a disk hot-spot interpretation is unlikely. 
Such a scenario also provides no natural explanation for the observed EPO.
Random flaring driven by internal jet processes, such as a sequence of intrinsic flares, could account for complex $\gamma$-ray variability \citep[e.g.,][]{Marscher1985ApJ298, Marscher2014ApJ780}, but does not naturally explain either the systematic increase of the EPO timescale or the contemporaneous two-dimensional motion of B10.
Shock interactions, particularly those involving stationary jet features, may contribute to individual flares \citep[e.g.,][]{Agudo2011ApJ735L, Fichet2022A&A661} and could be relevant to the final $\gamma$-ray peak.
Alternative kinematic interpretations of the VLBA component motion should also be considered.
A sequence of ballistic disturbances could in principle produce changes in Doppler boosting if different knots were successively ejected along different directions.
However, the five $\gamma$-ray enhancements are not accompanied by a corresponding sequence of newly emerging superluminal components in the VLBA observations.
This makes an interpretation in which each $\gamma$-ray enhancement is produced by a separate ballistic knot less natural.
Non-ballistic motion remains possible, with the helical trajectory adopted here representing one particular form of non-ballistic motion.

Combining the $\gamma$-ray EPO, the VLBA kinematic measurements, and the optical polarization observations, we consider helical jet motion to provide a geometric interpretation consistent with the observed variability.
In this scenario, knot B10 is treated as an emitting blob propagating downstream along a helical trajectory.
During its propagation, variations in the angle between the velocity of the emitting blob and the line of sight modulate the Doppler factor and thereby produce the observed $\gamma$-ray flux modulation.
Although additional radiative or kinematic processes may contribute to individual enhancements, the helical jet motion provides a possible framework for connecting the $\gamma$-ray EPO with the contemporaneous VLBA kinematics.

\subsection{Locating the inferred helical trajectory and its possible magnetic interpretation}

The evolution of knot B10 allows us to place model-dependent constraints on the location of the inferred helical trajectory associated with the EPO.
This estimation depends on the adopted coned jet geometry, de-projection angle, and core-shift measurement.
First of all, the distance between the observed 43 GHz VLBI core and the jet apex can be estimated through the core-shift effect.
The core shift between 14.4 and 43.1 GHz was measured by \citet{Marti-Vidal2016A&A596} to be $0.108\pm0.026$~mas.
The core shift in units of pc GHz is expressed as
\begin{equation}
    \Omega_{\rm rv} = 4.85 \times 10^{-9} \frac{\Delta r_{\rm core, \nu_{\rm 1}\nu_{\rm 2}} d_{\rm L}}{(1+{\rm z})^2} \times \frac{\nu_{\rm 1} \nu_{\rm 2}}{\nu_{\rm 2}-\nu_{\rm 1}}\,\, {\rm pc \ GHz},
\end{equation}
where $\Delta r_{\rm core,\nu_1\nu_2}$ is the measured core shift in mas, and $d_{\rm L}$ is the luminosity distance in pc.
The de-projected distance of the apparent VLBI core from the jet apex is then given by \citep{Lobanov1998A&A330}
\begin{equation}
    r_{\rm core}(\nu) = \frac{\Omega_{\rm rv}}{\nu \sin \psi},
\end{equation}
where $\nu$ is the observing frequency in GHz,
the model-derived  $\psi$ is the angle between the mean jet axis and the line of sight as the de-projection angle.
We obtained the 43 GHz VLBI core lies at a de-projected distance of $r_{\rm core}=6.38\pm1.89$~pc from the jet apex, based on the core-shift measurement of $\Delta r_{\rm core}=0.108\pm0.026$~mas between 14.4 and 43.1 GHz \citep{Marti-Vidal2016A&A596} and together with the model-derived angle $\psi=1.85^{\circ}\pm0.32^{\circ}$.

Then, knot B10 was first detected at a projected separation of $0.16\pm0.03$~mas from the 43 GHz core, corresponding to a de-projected distance of $19.18\pm4.59$~pc downstream of the core under the adopted jet geometry.
During its subsequent evolution, B10 advances by an additional projected distance of $0.38\pm0.09$~mas, corresponding to a further de-projected distance of $44.60\pm12.84$~pc.
In the end, the outermost portion of the trajectory is thus located at approximately $70.16\pm15.65$~pc from the jet apex.
Within the assumptions of the model, the inferred displacement of helical trajectory is therefore extended from approximately $25.56$ to $70.16$~pc from the jet apex.
The uncertainties were derived using standard error-propagation equation based on uncertainties of the measurements and model-dependent parameters.
We note that this distance estimate is highly model-dependent, particularly sensitive to the viewing angle $\psi$. In addition, the core-shift measurements were obtained at a different epoch (2010), which may introduce additional uncertainty into the inferred distance from the jet apex owing to intrinsic variability of the source over time.

The existence of helical knot motion trajectory suggests two possible scenarios regarding the jet configuration. 
In the first scenario, the conical jet propagates along a straight path, while the knot possesses both a velocity component parallel to the jet axis and a transverse velocity component perpendicular to the axis, possibly following a helical magnetic field structure.
In the second scenario, the jet itself may possess a helical geometry on these scales, causing the knot to follow a helical trajectory.
However, it remains challenging to distinguish between these two scenarios based solely on current observations. 
Future high-frequency VLBI observations, especially those capable of resolving the jet-launching region, together with long-term polarimetric monitoring, will provide critical constraints on the magnetic-field geometry and jet morphology.

Helical and filamentary structures have been observed in several relativistic jets on parsec and larger scales.
For example, high-resolution radio observations have revealed complex helical or filamentary morphologies in 3C~273, 3C~279, and M87 \citep{Lobanov2001Sci294, Fuentes2023NatAs7, Nikonov2023MNRAS526}.
However, the physical origin of such structures remains uncertain.
Such structures can be produced by different processes, including plasma instabilities in relativistic jets.
For instance, three-dimensional general relativistic magneto-hydrodynamic (GRMHD) simulations have shown that magnetized jets can develop current-driven kink instabilities, which may distort the jet morphology and generate helical structures \citep{Mignone2010MNRAS402, Blandford2019ARA&A57}.
Such instabilities may also provide conditions for particle acceleration through magnetic dissipation \citep{Alves2018PhRvL121}.
Then, these particles could radiate through interaction between magnetic field or photons and produce multi-wavelength emission.

Relativistic jets are generally expected to be launched through large-scale magnetic fields associated with the rotating black hole and/or the magnetized accretion flow, including the Blandford-Znajek (BZ) and Blandford-Payne (BP) mechanisms \citep{Blandford1977, Blandford1982}.
Previous population studies have suggested that the BZ mechanism may play an important role in powering BL Lac jets \citep[e.g.,][]{Xiao2022ApJ_1, Xie2024ApJ976}.
The helical trajectory inferred here is located tens of parsecs downstream of the jet apex, where the physical conditions may differ substantially from those in the jet-launching region.
This result therefore does not distinguish between the BZ and BP mechanisms, nor does it provide an independent constraint on the jet-launching process.
Thus, we only report a helical trajectory is found within 0.1 kpc of S5 0716+714 jet, based on the $\gamma$-ray EPO behaviour, VLBI observed kinematic motions and optical polarization observation.
And, our analysis and model provides a framework for connecting parsec-scale jet evolution and the observed $\gamma$-ray EPO through geometric Doppler modulation.

\subsection{Conclusions}

In this work, we investigate the $\gamma$-ray variability of S5 0716+714 from July 2014 to June 2015 and explore its connection with the parsec-scale jet evolution. 
The adaptive-binning \textit{Fermi}-LAT light curve shows five consecutive flux enhancements with progressively increasing amplitudes and peak separations, referred to as EPO.
And, the WWZ analysis was applied to the 3-day-binned $\gamma$-ray light curves further confirms this evolution of the increasing characteristic period, which distinguishes the EPO from typical stationary periodic signals.
A local significance of $3.26\sigma$ and a global significance of $2.60$--$2.64\sigma$ is achieved for the EPO signal.
The temporal association between the EPO and the parsec-scale jet evolution suggests a possible connection between $\gamma$-ray variability and the motion of knot B10. 
A helical trajectory model is employed and successfully explains both the VLBI-observed knot motion and the $\gamma$-ray EPO. 
Combining the model with the core-shift measurement, we find that knot B10 propagated over a distance of $44.60\pm12.84$~pc and reached a location of $70.16\pm15.65$~pc from the jet apex. 
This work provides a framework for investigating the connection between parsec-scale jet evolution and high-energy variability in relativistic jets.

\section*{Acknowledgements}
H.B.X acknowledges the support from the National Natural Science Foundation of China (NSFC 12203034), the Shanghai Science and Technology Fund (22YF1431500), the science research grants from the China Manned Space Project (CMS-CSST-2025-A07), and the Shanghai Municipal Education Commission regarding artificial intelligence empowered research.
Z.J.L acknowledges the support from the Shanghai Science and Technology Foundation Fund under grant No. 20070502400, and the science research grants from the China Manned Space Project. 
S.H.Z acknowledges support from the National Natural Science Foundation of China (grant No. 12173026), the Program for Professor of Special Appointment (Eastern Scholar) at Shanghai Institutions of Higher Learning and the Shuguang Program (23SG39) of the Shanghai Education Development Foundation and Shanghai Municipal Education Commission.
M.M acknowledges the Croatian Science Foundation (HrZZ) under the project number IP-2022-10-4595 and the University of Rijeka Project uniri-mzi-25-3 funded by the European Union - NextGenerationEU.
L.C.H was supported by the National Science Foundation of China (12233001) and the China Manned Space Program (CMS-CSST-2025-A09).
J.H.F acknowledges the support from the National Natural Science Foundation of China (NSFC 12433004, U2031201), the Eighteenth Regular Meeting Exchange Project of the Scientific and Technological Cooperation Committee between the People's Republic of China and the Republic of Bulgaria (Series No. 1802), and the science research grants from the China Manned Space Project with  NO. CMS-CSST-2025-A07.

This study makes use of VLBA data from the VLBA-BU Blazar Monitoring Program (BEAM-ME and VLBA-BU-BLAZAR;
http://www.bu.edu/blazars/BEAM-ME.html), funded by NASA through the Fermi Guest Investigator Program. The VLBA is an instrument of the National Radio Astronomy Observatory. The National Radio Astronomy Observatory is a facility of the National Science Foundation operated by Associated Universities, Inc.

\appendix

\section{Power spectral density fitting and probability distribution function fitting}\label{appendix:PSD_PDF}
\setcounter{figure}{0}
\setcounter{table}{0}

To construct the stochastic null hypothesis used in the Monte Carlo significance analysis, we characterized both the power spectral density (PSD) and the flux probability density function (PDF) of the 3-day-binned $\gamma$-ray light curve over MJD 56840 -57200.

\begin{figure*}
\centering
\includegraphics[width=0.85\textwidth]{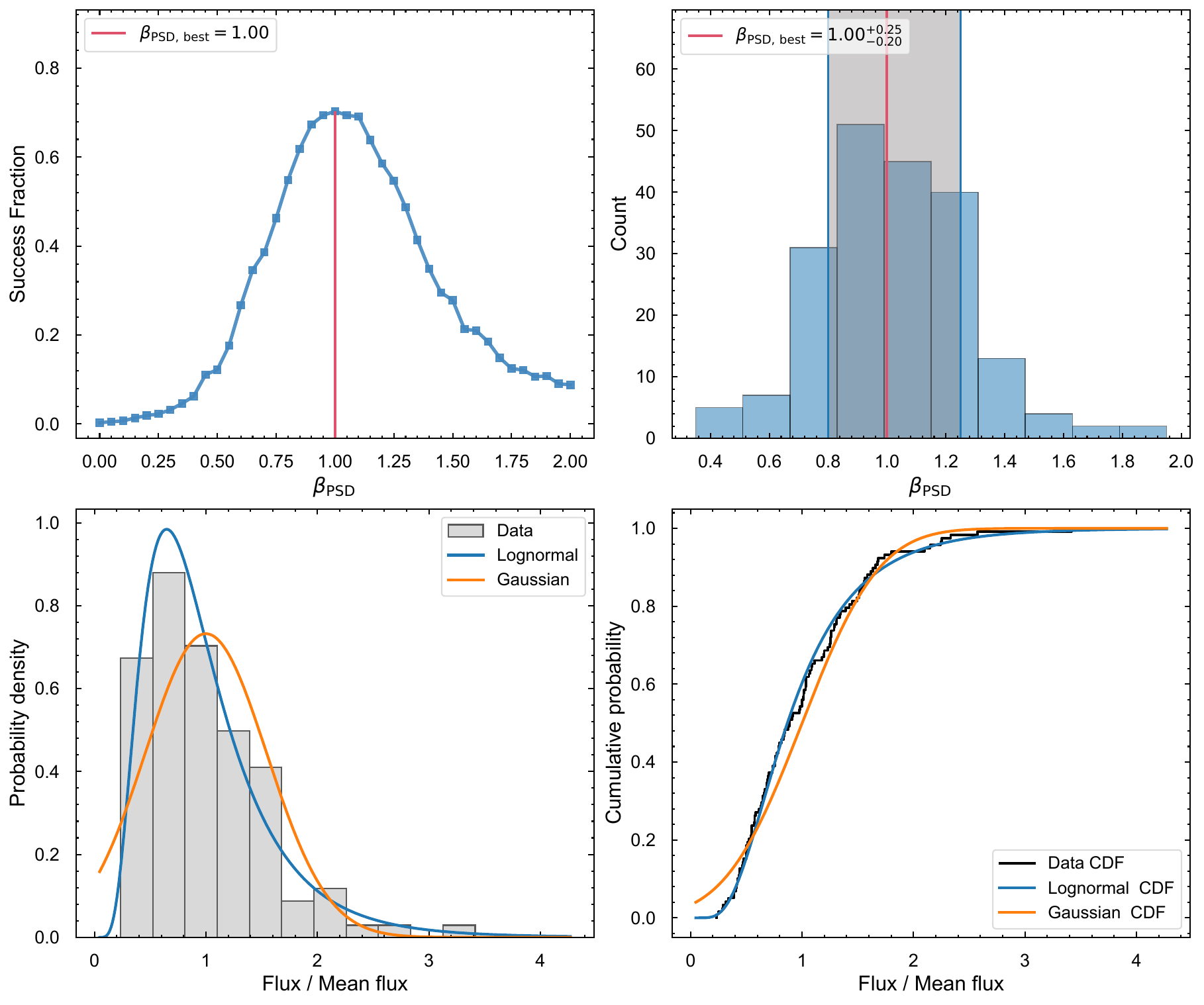}
\caption{
Power spectral density and probability distribution function fitting of the 3-day-binned $\gamma$-ray light curve.
Upper-left panel: PSRESP success fraction as a function of the trial PSD slope $\beta$. 
The maximum success fraction occurs at $\beta_{\rm PSD}=1.0$.
Upper-right panel: Histogram of the PSD slopes obtained from 200 simulated light curves generated with $\beta_{\rm PSD}=1.0$.
Lower-left panel: Probability density of the normalized observed fluxes and the fitted Gaussian (orange) and lognormal (blue) functions.
Lower-right panel: Empirical and fitted cumulative distribution functions.
}
\label{fig:PDF_and_PDF}
\end{figure*}

We modeled the PSD of the observed light curve as a power law $P(f)\propto f^{-\beta_{\rm PSD}}$, where $\beta_{\rm PSD}$ is the PSD slope.
We estimated the PSD slope through a forward-folding method using the power spectral response method \citep[PSRESP;][]{Uttley2002MNRAS332, Chatterjee2008ApJ689, Max-Moerbeck2014MNRAS445}.
Specifically, we examined trial slopes from $\beta=0$ to $2.00$ in steps of 0.05. For each trial slope value, $N_{\rm PSRESP}=3000$ artificial light curves were generated with the same temporal sampling as the observed light curve and with durations 100 times longer than that of the observed light curve. 
For each trial slope, we calculated two $\chi^2$-like functions, which compared the discrepancy between the observed PSD and the simulated PSDs, 
\begin{equation}
    \chi_{\rm obs}^{2} = \sum_{} \frac{ \left( {\rm PSD}_{\rm obs} - \overline{{\rm PSD}}_{\rm sim} \right)^2 }{ (\Delta {\rm PSD}_{\rm sim})^{2} },
\end{equation}
and
\begin{equation}
    \chi_{{\rm dist},i}^{2} = \sum_{} \frac{ \left( {\rm PSD}_{{\rm sim},i} - \overline{{\rm PSD}}_{\rm sim} \right)^2 }{ (\Delta {\rm PSD}_{\rm sim})^{2} },
\end{equation}
where $\overline{{\rm PSD}}_{\rm sim}$ is the average of ${\rm PSD}_{{\rm sim},i}$ and $\Delta {\rm PSD}_{\rm sim}$ is the standard deviation of the ${\rm PSD}_{{\rm sim},i}$.
The goodness of fit was described by ``success fraction'', defined as
\begin{equation}
    {\rm Success~Fraction} = \frac{1}{N_{\rm PSRESP}} \sum_{i=1}^{N_{\rm PSRESP}} I \left( \chi_{{\rm dist},i}^{2} > \chi_{\rm obs}^{2} \right),
\end{equation}
where $I$ is the indicator function.
The upper-left panel of Figure~\ref{fig:PDF_and_PDF} shows the success fraction as a function of the trial PSD slope, yielding a best-fit value of $\beta_{\rm PSD}=1.0$.
Following \citet{MAGIC2021A&A655} and \citet{MAGIC2025A&A694}, we estimated the uncertainty on the PSD slope by generating 200 simulated light curves using the best-fit PSD slope and applying the same fitting procedure as that used for the observed light curve. 
The resulting distribution of the recovered PSD slopes was used to determine the uncertainty, defined by the 68\% confidence interval. This distribution is shown in the upper-right panel of Figure~\ref{fig:PDF_and_PDF}. 
We therefore obtained a best-fit PSD slope of $\beta_{\rm PSD}=1.0^{+0.25}_{-0.20}$, which was subsequently adopted in the simulations in Section~\ref{sec:EPO_significance}.

The lower-left panel of Figure~\ref{fig:PDF_and_PDF} shows the distribution of the observed fluxes normalized by their mean value, $x=F/\langle F\rangle$. 
We fitted the normalized flux distribution with Gaussian and lognormal models. 
Their probability density functions are given by
\begin{equation}
    p_{\rm G}(x)= \frac{1}{\sqrt{2\pi}\sigma_{\rm G}} \exp\left[ -\frac{(x-\mu_{\rm G})^2}{2\sigma_{\rm G}^2} \right],
\end{equation}
and
\begin{equation}
p_{\rm LN}(x)= \frac{1}{x\sqrt{2\pi}\sigma_{\rm LN}} \exp\left[ -\frac{(\ln x-\mu_{\rm LN})^2} {2\sigma_{\rm LN}^2} \right],
\end{equation}
where $\mu_{\rm G}$ and $\sigma_{\rm G}$ are the mean and standard deviation of the Gaussian distribution, while $\mu_{\rm LN}$ and $\sigma_{\rm LN}$ are the mean and standard deviation of $\ln x$ for the lognormal distribution.
The goodness of fit was compared using the AIC as in Equation~\eqref{eq:aic}, in which $\mathcal{L}_{\rm max}$ is the maximum likelihood of the fitted Gaussian or lognormal distribution.  
The best-fitting parameters and AIC values are listed in Table~\ref{tab:PDF_fit}. 
The lognormal model has a lower AIC than the Gaussian model, indicating that it provides a better description of the observed flux distribution.
The lower-right panel of Figure~\ref{fig:PDF_and_PDF} compares the empirical cumulative distribution function (CDF) of the observed fluxes with the CDFs predicted by the two fitted models. 
The lognormal CDF follows the observed cumulative distribution more closely than the Gaussian CDF, further supporting the lognormal model as the preferred description of the observed flux distribution. 
We therefore adopted the best-fitting lognormal distribution when generating the artificial light curves.

\begin{table}
\centering
\caption{Best-fitting parameters and information criteria for the Gaussian and lognormal models of the normalized flux distribution.}
\label{tab:PDF_fit}
\begin{tabular}{lcccc}
\toprule
Model & $\mu_{\rm G}$ or $\mu_{\rm LN}$ & $\sigma_{\rm G}$ or $\sigma_{\rm LN}$ & $\ln\mathcal{L}_{\rm max}$  & AIC\\
\midrule
Gaussian & 1.00 & 0.54 & -95.76 & 195.52\\

Lognormal & -0.14 & 0.54 & -78.11 & 160.22\\
\bottomrule
\end{tabular}
\end{table}

\section{VLBA fitting component parameters}\label{appendix:VLBA_fitting_parameter}
\setcounter{figure}{0}
\setcounter{table}{0}

Table~\ref{Tab:VLBA_knot_params} lists the parameters of the VLBA jet components derived at each observing epoch.
The table includes the flux density, radial distance, position angle, angular size, and observed brightness temperature, along with their associated uncertainties.

\begin{sidewaystable*}
\centering
\footnotesize
\renewcommand{\arraystretch}{1.2}
\setlength{\tabcolsep}{5pt}
\caption{Time evolution of the VLBA jet component parameters in S5 0716+714.
For each component at each epoch, we report the flux density ($S$), radial distance from the core ($r$), position angle (\textit{PA}), angular size ($a$), observed brightness temperature ($T_{\rm b,~obs}$, where ``L'' denotes a lower limit), and position relative to the core (relative \textit{RA} and \textit{Dec}) for each component.
}\label{Tab:VLBA_knot_params}
\begin{tabular}{cccccccccccc}
    \toprule
    Epoch & Time & $\chi^2_{\rm red}$ & $S$ & $r$ & \textit{PA} & $a$ &
    $T_{\rm b,~obs}$ & Flag $T_{\rm b,~obs}$ & Rel. \textit{RA} & Rel. \textit{Dec} & Knot \\
     & (MJD) & & (Jy) & (mas) & (deg) & (mas) & (K) & &
    (mas) & (mas) & \\
    \midrule
    2014.570 & 56866 & 0.890
    & $1.616 \pm 0.081$
    & 0
    & 0
    & $0.020 \pm 0.005$
    & $3.03\times10^{12}$
    & L
    & $0.000 \pm 0.005$
    & $0.000 \pm 0.010$
    & A0 \\

    &
    &
    & $0.566 \pm 0.029$
    & $0.093 \pm 0.005$
    & $86.1 \pm 3.2$
    & $0.034 \pm 0.008$
    & $3.57\times10^{11}$
    &
    & $0.093 \pm 0.005$
    & $0.006 \pm 0.010$
    & A1 \\

    &
    &
    & $0.088 \pm 0.010$
    & $0.163 \pm 0.027$
    & $52.0 \pm 5.6$
    & $0.107 \pm 0.024$
    & $5.76\times10^{9}$
    &
    & $0.129 \pm 0.019$
    & $0.101 \pm 0.038$
    & B10 \\
    \midrule
    
    2014.726 & 56923 & 0.664
    & $1.322 \pm 0.066$
    & 0
    & 0
    & $0.020 \pm 0.005$
    & $2.45\times10^{12}$
    &
    & $0.000 \pm 0.005$
    & $0.000 \pm 0.010$
    & A0 \\

    &
    &
    & $0.205 \pm 0.012$
    & $0.097 \pm 0.006$
    & $77.6 \pm 3.2$
    & $0.031 \pm 0.010$
    & $1.60\times10^{11}$
    &
    & $0.095 \pm 0.006$
    & $0.021 \pm 0.011$
    & A1 \\

    &
    &
    & $0.241 \pm 0.015$
    & $0.187 \pm 0.011$
    & $69.1 \pm 2.8$
    & $0.093 \pm 0.017$
    & $2.09\times10^{10}$
    &
    & $0.174 \pm 0.010$
    & $0.066 \pm 0.019$
    & B10 \\

    &
    &
    & $0.033 \pm 0.012$
    & $0.484 \pm 0.124$
    & $19.1 \pm 4.4$
    & $0.189 \pm 0.040$
    & $6.89\times10^{8}$
    &
    & $0.158 \pm 0.065$
    & $0.457 \pm 0.130$
    & A2 \\
    \midrule
    
    2014.871 & 56976 & 0.369
    & $2.658 \pm 0.133$
    & 0
    & 0
    & $0.019 \pm 0.004$
    & $4.98\times10^{12}$
    & L
    & $0.000 \pm 0.005$
    & $0.000 \pm 0.010$
    & A0 \\

    &
    &
    & $0.052 \pm 0.008$
    & $0.124 \pm 0.007$
    & $66.2 \pm 2.6$
    & $0.000 \pm 0.012$
    & $9.82\times10^{10}$
    & L
    & $0.114 \pm 0.006$
    & $0.050 \pm 0.012$
    & A1 \\

    &
    &
    & $0.272 \pm 0.016$
    & $0.212 \pm 0.015$
    & $61.4 \pm 2.8$
    & $0.116 \pm 0.019$
    & $1.52\times10^{10}$
    &
    & $0.186 \pm 0.011$
    & $0.102 \pm 0.023$
    & B10 \\
    \midrule
    
    2014.992 & 57020 & 0.795
    & $1.167 \pm 0.059$
    & 0
    & 0
    & $0.016 \pm 0.005$
    & $2.19\times10^{12}$
    & L
    & $0.000 \pm 0.005$
    & $0.000 \pm 0.010$
    & A0 \\

    &
    &
    & $0.169 \pm 0.011$
    & $0.118 \pm 0.008$
    & $50.8 \pm 2.2$
    & $0.025 \pm 0.010$
    & $2.00\times10^{11}$
    &
    & $0.092 \pm 0.005$
    & $0.075 \pm 0.011$
    & A1 \\

    &
    &
    & $0.256 \pm 0.016$
    & $0.308 \pm 0.020$
    & $54.0 \pm 2.2$
    & $0.138 \pm 0.021$
    & $1.01\times10^{10}$
    &
    & $0.249 \pm 0.014$
    & $0.181 \pm 0.028$
    & B10 \\
    \midrule
    
    2015.121 & 57067 & 1.026
    & $1.980 \pm 0.099$
    & 0
    & 0
    & $0.018 \pm 0.005$
    & $3.71\times10^{12}$
    & L
    & $0.000 \pm 0.005$
    & $0.000 \pm 0.010$
    & A0 \\

    &
    &
    & $0.247 \pm 0.014$
    & $0.147 \pm 0.010$
    & $51.2 \pm 2.2$
    & $0.055 \pm 0.013$
    & $6.15\times10^{10}$
    &
    & $0.115 \pm 0.007$
    & $0.092 \pm 0.013$
    & A1 \\

    &
    &
    & $0.138 \pm 0.013$
    & $0.349 \pm 0.061$
    & $41.0 \pm 4.6$
    & $0.241 \pm 0.032$
    & $1.78\times10^{9}$
    &
    & $0.229 \pm 0.037$
    & $0.264 \pm 0.074$
    & B10 \\
    \midrule
    
    2015.274 & 57123 & 0.465
    & $2.094 \pm 0.105$
    & 0
    & 0
    & $0.017 \pm 0.005$
    & $3.93\times10^{12}$
    & L
    & $0.000 \pm 0.005$
    & $0.000 \pm 0.010$
    & A0 \\

    &
    &
    & $0.373 \pm 0.020$
    & $0.089 \pm 0.009$
    & $51.3 \pm 3.3$
    & $0.057 \pm 0.012$
    & $8.55\times10^{10}$
    &
    & $0.069 \pm 0.006$
    & $0.056 \pm 0.012$
    & A1 \\

    &
    &
    & $0.045 \pm 0.010$
    & $0.235 \pm 0.041$
    & $44.1 \pm 4.9$
    & $0.101 \pm 0.027$
    & $3.34\times10^{9}$
    &
    & $0.163 \pm 0.026$
    & $0.169 \pm 0.051$
    & T10 \\

    &
    &
    & $0.089 \pm 0.011$
    & $0.484 \pm 0.058$
    & $36.3 \pm 2.9$
    & $0.179 \pm 0.030$
    & $2.09\times10^{9}$
    &
    & $0.287 \pm 0.034$
    & $0.390 \pm 0.067$
    & B10 \\
    \midrule
    
    2015.356 & 57153 & 0.481
    & $1.581 \pm 0.079$
    & 0
    & 0
    & $0.023 \pm 0.005$
    & $2.29\times10^{12}$
    &
    & $0.000 \pm 0.005$
    & $0.000 \pm 0.010$
    & A0 \\

    &
    &
    & $0.309 \pm 0.017$
    & $0.098 \pm 0.010$
    & $48.7 \pm 3.2$
    & $0.063 \pm 0.013$
    & $5.92\times10^{10}$
    &
    & $0.074 \pm 0.007$
    & $0.065 \pm 0.013$
    & A1 \\

    &
    &
    & $0.032 \pm 0.010$
    & $0.262 \pm 0.043$
    & $50.9 \pm 5.4$
    & $0.095 \pm 0.029$
    & $2.65\times10^{9}$
    &
    & $0.204 \pm 0.029$
    & $0.165 \pm 0.059$
    & T10 \\

    &
    &
    & $0.069 \pm 0.012$
    & $0.498 \pm 0.078$
    & $35.1 \pm 3.7$
    & $0.201 \pm 0.034$
    & $1.28\times10^{9}$
    &
    & $0.286 \pm 0.045$
    & $0.407 \pm 0.090$
    & B10 \\
    \midrule
    
    2015.436 & 57182 & 0.884
    & $0.947 \pm 0.048$
    & 0
    & 0
    & $0.018 \pm 0.006$
    & $1.78\times10^{12}$
    & L
    & $0.000 \pm 0.005$
    & $0.000 \pm 0.010$
    & A0 \\

    &
    &
    & $0.308 \pm 0.017$
    & $0.144 \pm 0.011$
    & $47.5 \pm 2.2$
    & $0.067 \pm 0.014$
    & $5.18\times10^{10}$
    &
    & $0.106 \pm 0.007$
    & $0.098 \pm 0.014$
    & A1 \\

    &
    &
    & $0.039 \pm 0.010$
    & $0.298 \pm 0.027$
    & $50.3 \pm 2.9$
    & $0.069 \pm 0.023$
    & $6.13\times10^{9}$
    &
    & $0.229 \pm 0.018$
    & $0.191 \pm 0.036$
    & T10 \\

    &
    &
    & $0.054 \pm 0.012$
    & $0.542 \pm 0.083$
    & $36.2 \pm 3.6$
    & $0.189 \pm 0.035$
    & $1.14\times10^{9}$
    &
    & $0.320 \pm 0.048$
    & $0.437 \pm 0.096$
    & B10 \\
    
    \bottomrule
\end{tabular}
\end{sidewaystable*}

\section{MCMC Posterior Distributions and Convergence Diagnostics}
\label{appendix:mcmc_convergence}

\setcounter{figure}{0}
\setcounter{table}{0}

\begin{figure*}
\centering
\includegraphics[width=0.85\textwidth]{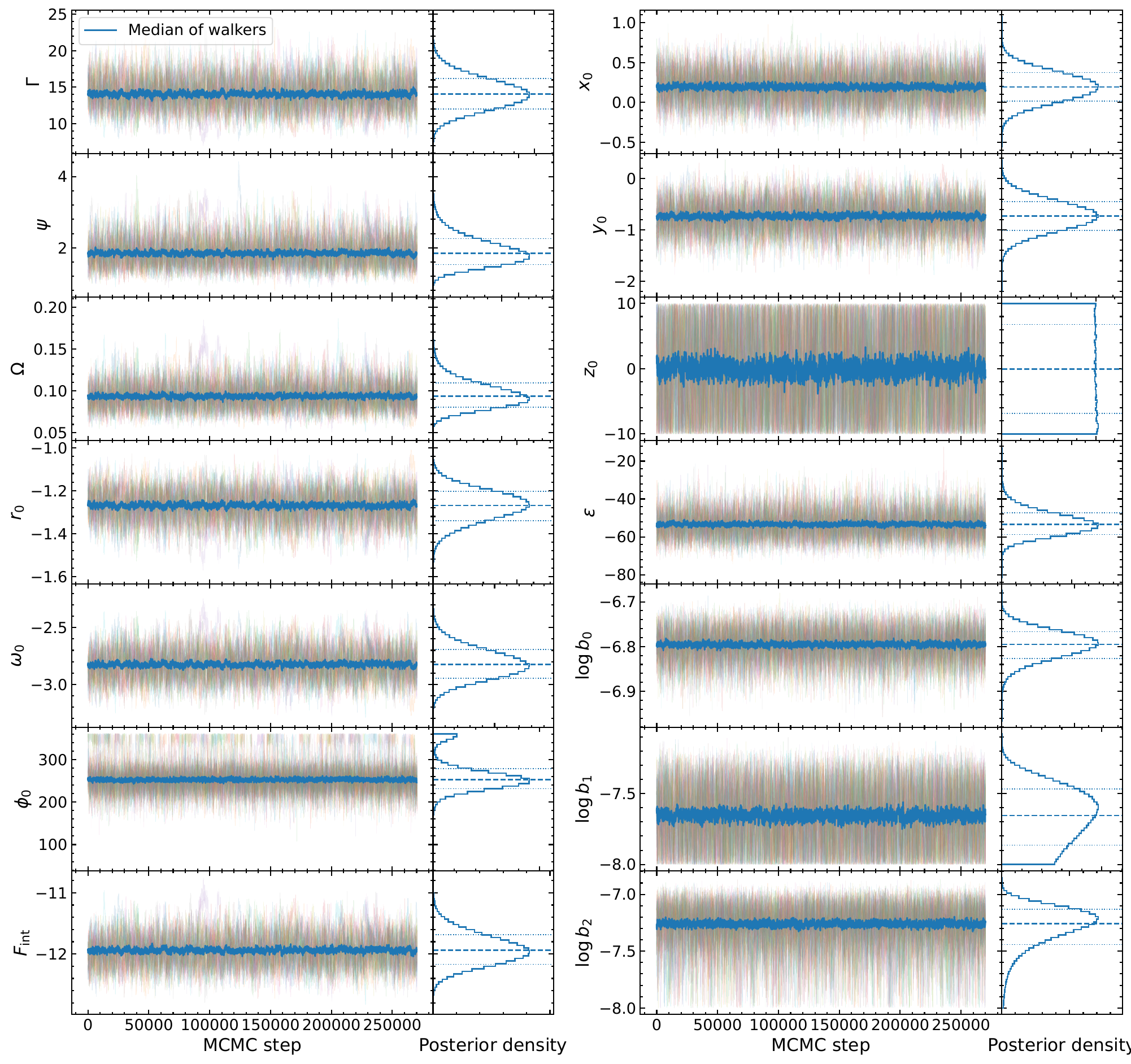}
\caption{
For each parameter, the left panel shows the MCMC traces for a representative subset of walkers, along with the median for all walkers, while the right panel presents the corresponding marginalized posterior distribution.
}
\label{fig:mcmc_convergence}
\end{figure*}
\begin{figure*}
\centering
\includegraphics[width=0.95\textwidth]{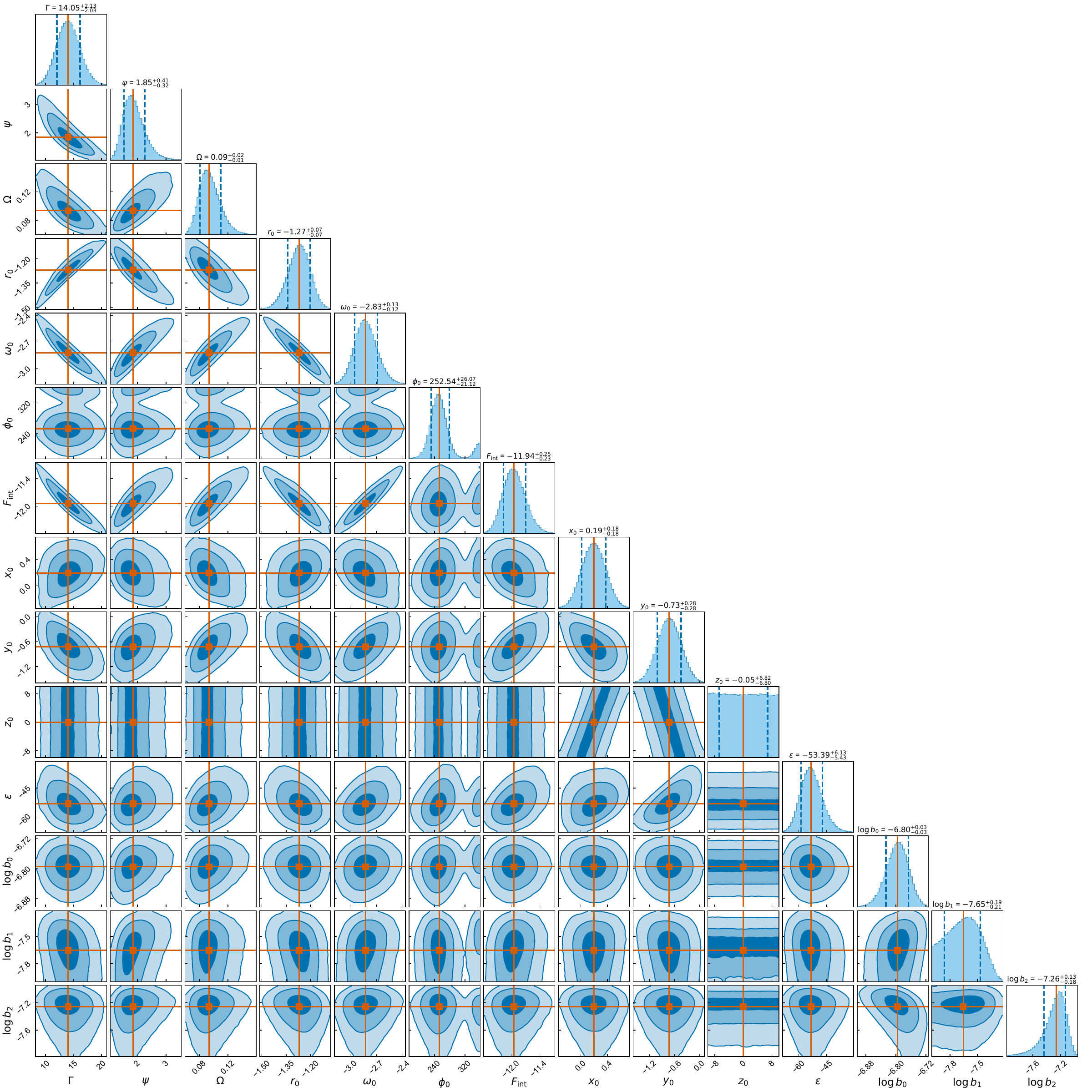}
\caption{
Corner plot showing the results of the MCMC analysis for the fitted helical model.
The histograms present the marginalized posterior distributions for each parameter.
The best-fit values are taken from the 50th percentiles, as shown by the red solid lines, and the associated uncertainties correspond to the 16th and 84th percentiles. 
}
\label{fig:mcmc}
\end{figure*}

Figure~\ref{fig:mcmc_convergence} presents the corresponding trace plots and marginalized posterior distributions for all model parameters. 
The trace plots illustrate the evolution of the walkers after the burn-in phase, together with the median across all walkers, while the adjacent panels show the sampled posterior distributions.
Figure~\ref{fig:mcmc} shows the posterior distributions of the fitted helical model parameters derived from the MCMC analysis.

\bibliographystyle{elsarticle-harv} 
\bibliography{lib_xiao}






\end{document}